\documentclass[journal]{IEEEtran}
\usepackage[compress]{cite}
\usepackage{amsmath}
\usepackage{amssymb}
\usepackage{mathrsfs}
\usepackage{enumerate}
\usepackage{tikz}
\usepackage{adjustbox}
\usepackage{xcolor}
\usetikzlibrary{shapes,arrows,chains}
\usepackage{graphicx}
\usepackage{algorithm}
\usepackage{algpseudocode}
\usepackage{xpatch}
\usepackage{mathtools}
\usepackage{nicefrac}
\usepackage{float}
\floatstyle{plaintop}
\restylefloat{table}
\usepackage{cleveref}
\usepackage{resizegather}
\DeclarePairedDelimiter{\norm}{\lVert}{\rVert}
\algdef{SE}[DOWHILE]{Do}{doWhile}{\algorithmicdo}[1]{\algorithmicwhile\ #1}%
\algrenewcommand\alglinenumber[1]{{\sffamily\footnotesize#1}}
\makeatletter
\xpatchcmd{\algorithmic}{\itemsep\z@}{\itemsep=.25ex plus2pt}{}{}
\makeatother

\usepackage{color}
\usepackage{soul}
\usepackage{environ}         
\usepackage{etoolbox}        
\usepackage[acronym]{glossaries}
\newacronym{1g}{1G}{first-generation}
\newacronym{4g}{4G}{fourth-generation}
\newacronym{5g}{5G}{fifth-generation}
\newacronym{mimo}{MIMO}{multiple-input-multiple-output}
\newacronym{ris}{RIS}{reconfigurable inteligent surface}
\newacronym{siso}{SISO}{single-input-single-output}
\newacronym{mamimo}{MaMIMO}{massive multiple-input-multiple-output}
\newacronym{sumimo}{SU-MIMO}{single user MIMO}
\newacronym{mumimo}{MU-MIMO}{multi user MIMO}
\newacronym{embms}{eMBMS}{evolved Multimedia Broadcast and Multicast Service}
\newacronym{sca}{SCA}{successive convex approximation}
\newacronym{sinr}{SINR}{signal-to-interference-plus-noise ratio}
\newacronym{ula}{ULA}{uniform linear array}
\newacronym{mcs}{MCS}{modulation and coding scheme}
\newacronym{mrt}{MRT}{maximum ratio transmission}
\newacronym{zf}{ZF}{zero-forcing}
\newacronym{mr}{MR}{maximum ratio}
\newacronym{se}{SE}{spectral efficiency}
\newacronym{sse}{SumSE}{sum spectral efficiency}
\newacronym{mise}{MinSE}{minimum spectral efficiency}
\newacronym{asd}{ASD}{angular standard deviation}
\newacronym{adr}{ADR}{aggregated data rate}
\newacronym{embb}{eMBB}{enhanced mobile broadband}
\newacronym{mmtc}{mMTC}{massive machine type communications}
\newacronym{urllc}{URLLC}{ultra reliable low latency communications}
\newacronym{csi}{CSI}{channel state information}
\newacronym{pmi}{PMI}{precoding matrix indicator}
\newacronym{ri}{RI}{rank indicator}
\newacronym{csi-rs}{CSI-RS}{CSI-reference signal}
\newacronym{cri}{CRI}{CSI-RS resource indicator}
\newacronym{bs}{BS}{base station}
\newacronym{re}{RE}{resource element}
\newacronym{mmwave}{mmWave}{millimeter-wave}
\newacronym{umwave}{$\mu$mWaves}{micrometer waves}
\newacronym{rnn}{RNN}{recurrent neural network}
\newacronym{cnn}{CNN}{convolutional neural network}
\newacronym{ngmn}{NGMN}{next-generation mobile network}
\newacronym{lte}{LTE}{Long Term Evolution}
\newacronym{lte-a}{LTE-A}{Long Term Evolution Advanced}
\newacronym{5gnr}{5G NR}{5G New Radio}
\newacronym{mm}{MM}{mixed mode}
\newacronym{cdf}{CDF}{cumulative distribution function}
\newacronym{phy}{PHY}{physical}
\newacronym{mac}{MAC}{medium access control}
\newacronym{3gpp}{3GPP}{3rd Generation Partnership Project}
\newacronym{fdd}{FDD}{frequency division duplexing}
\newacronym{tdd}{TDD}{time division duplexing}
\newacronym{ofdm}{OFDM}{orthogonal frequency division multiplexing}
\newacronym{ss}{SS}{synchronization signal} 
\newacronym{pss}{PSS}{primary synchronization signal} 
\newacronym{sss}{SSS}{secondary synchronization signal} 
\newacronym{pbch}{PBCH}{physical broadcast channel} 
\newacronym{dmrs}{DMRS}{demodulation reference signal} 
\newacronym{gnb}{gNB}{next generation nodeB} 
\newacronym{rsrp}{RSRP}{reference signal received power} 
\newacronym{rrm}{RRM}{radio resource management} 
\newacronym{srs}{SRS}{sounding reference signal} 
\newacronym{ran}{RAN}{radio access network} 
\newacronym{nn}{NN}{neural network} 
\newacronym{ue}{UE}{user equipment} 
\newacronym{awgn}{AWGN}{additive white Gaussian noise} 
\newacronym{epa}{EPA}{Extended Pedestrian A model}
\newacronym{eva}{EVA}{Extended Vehicular A model}
\newacronym{etu}{ETU}{Extended Typical Urban model}
\newacronym{tdl}{TDL}{tapped delay line}
\newacronym{cdl}{CDL}{clustered delay line}
\newacronym{uma}{UMa}{urban macro-cell}
\newacronym{isd}{ISD}{inter-site distance}
\newacronym{nlos}{NLOS}{non-line of sight}
\newacronym{los}{LOS}{line of sight}
\newacronym{o2o}{O2O}{outdoor-to-outdoor}
\newacronym{o2i}{O2I}{outdoor-to-indoor}
\newacronym{ul}{UL}{uplink}
\newacronym{dl}{DL}{downlink}
\newacronym{ls}{LS}{least squares}
\newacronym{mmse}{MMSE}{minimum mean square error}
\newacronym{snr}{SNR}{signal-to-noise ratio}
\newacronym{mse}{MSE}{mean square error}
\newacronym{nr}{NR}{New Radio}
\newacronym{prb}{PRB}{physical resource block}
\newacronym{scs}{SCS}{subcarrier spacing}
\newacronym{bler}{BLER}{block error rate}
\newacronym{smmmra}{SMMMRA}{subgroup multicast \gls{mamimo} resource allocation}
\newacronym{mmf}{MMF}{max-min fairness}
\newacronym{smmu}{SMMU}{subgroups of multicast \gls{mamimo} users}
\newacronym{gsmma}{GSMMA}{greedy subgroup multicast \gls{mamimo} algorithm}
\include{cite}

\newcommand{\herm}{^\mathsf{H}}
\newcommand{\trans}{^\mathsf{T}}
\newcommand{\taup}{\tau_{\mathrm{p}}}
\newcommand{\UL}{^{\mathrm{ul}}}

\ifCLASSOPTIONcompsoc
    \usepackage[caption=false, font=normalsize, labelfont=sf, textfont=sf, subrefformat=parens, labelformat=parens]{subfig}
\else
\usepackage[caption=false, font=footnotesize, subrefformat=parens, labelformat=parens]{subfig}
\fi

\begin{document}
\title{User Subgrouping and Power Control for Multicast Massive MIMO over Spatially Correlated Channels}

\author{Alejandro de la Fuente, Giovanni Interdonato~\IEEEmembership{Member,~IEEE}, and Giuseppe Araniti,~\IEEEmembership{Senior~Member,~IEEE} 

\thanks{A. de la Fuente is with the Department of Signal Theory and Communications, Universidad Rey Juan Carlos, Camino del Molino, 28943, Fuenlabrada (Madrid), Spain (email: alejandro.fuente@urjc.es).

G. Interdonato is with the Department of Electrical and Information Engineering, University of Cassino and Southern Latium, Cassino, Italy (e-mail: giovanni.interdonato@unicas.it).

G. Araniti is with the  Department  of  Information  Engineering,  Infrastructure  and  Sustainable Energy (DIIES), University Mediterranea of Reggio Calabria, Reggio Calabria 89060, Italy. (email: araniti@unirc.it).}

\thanks{An excerpt of this article has been published in the proceedings of the 2021 IEEE International Conference on Communications (ICC)~\cite{2021delaFuente}.}

\thanks{The work of A.~de~la~Fuente has been supported by the Spanish Ministry of Education and Professional Training Program “José Castillejo 2019-2020”. The work of G.~Interdonato has been supported by the Italian Ministry of Education University and Research (MIUR) PRIN 2017 Project “LiquidEdge”.}
}

\markboth{IEEE Transactions on Broadcasting}{Alejandro de la Fuente \MakeLowercase{\text it{et al.}}: User Subgrouping and Resource Allocation in Multicast Massive MIMO over Spatially Correlated Rayleigh Fading Channels.}

\maketitle

\begin{abstract}
Massive \gls{mimo} is unquestionably a key enabler of the fifth-generation (5G) technology for mobile systems, enabling to meet the high requirements of upcoming mobile broadband services. Physical-layer multicasting refers to a technique for simultaneously serving multiple users, demanding for the same service and sharing the same radio resources, with a single transmission. 
Massive \gls{mimo} systems with multicast communications have been so far studied under the ideal assumption of uncorrelated Rayleigh fading channels. In this work, we consider a practical multicast massive \gls{mimo} system over spatially correlated Rayleigh fading channels, investigating the impact of the spatial channel correlation on the \emph{favorable propagation}, hence on the performance. We propose a subgrouping strategy for the multicast users based on their channel correlation matrices' similarities. The proposed subgrouping approach capitalizes on the spatial correlation to enhance the quality of the channel estimation, and thereby the effectiveness of the precoding. Moreover, we devise a \gls{mmf} power allocation strategy that makes the \gls{se} among different multicast subgroups uniform. Lastly, we propose a novel power allocation for \gls{ul} pilot transmission to maximize the \gls{se} among the users within the same multicast subgroup. Simulation results show a significant \gls{se} gain provided by our user subgrouping and power allocation strategies. Importantly, we show how spatial channel correlation can be exploited to enhance multicast massive \gls{mimo} communications.   
\end{abstract}

\glsresetall
\section{Introduction}

\IEEEPARstart{T}{he} \gls{5g} technology standard for wireless cellular systems has established high performance requirements for the upcoming mobile services, in terms of data rates, latency, and number of connected devices \cite{2019Ericsson}. Massive \gls{mimo} is a physical layer technology that makes use of multiple antennas at the \gls{bs} to jointly and coherently serve multiple users in the same time-frequency resources~\cite{2016Marzetta,2017Bjornsonbook}. By providing extraordinary levels of array gain, spatial multiplexing gain, and spatial diversity, massive \gls{mimo} is able to meet the 5G requirements~\cite{2014Andrews,2014Boccardi}. 
%
Massive MIMO is not a promising key-enabling technology for 5G any longer but became reality.  Many aspects, techniques, methods and protocols of massive MIMO are part of the “New Radio” (i.e., the commercial name of 5G) standard. Industrial research has acknowledged firstly, and corroborated on field lately, the accuracy and the value of the academic theoretical research on massive MIMO. Undoubtedly, massive MIMO technology will become the foundation for beyond 5G networks, where new applications and designs of antenna arrays are already on the table. Research in this direction is ongoing under the names of cell-free massive MIMO \cite{Interdonato2019}, holographic MIMO \cite{Huang2020}, large intelligent surface \cite{Hu2018} and intelligent reflecting surface (IRS) \cite{Wu2019}, etc. 
We refer interested readers to \cite{Bjornson2019}, \cite{Zhang2020} and references therein for a comprehensive overview on the role of massive MIMO in beyond 5G networks.    
The intrinsic characteristics of massive \gls{mimo} systems operating at the sub-6 GHz frequency bands have been deeply studied.
In most propagation environments, massive \gls{mimo} offers two fundamental properties known as \emph{favorable propagation} and \emph{channel hardening}: as the number of \gls{bs} antennas increases, users' channels become nearly pairwise orthogonal and deterministic, respectively. These phenomena lead to a significant increase in spectral and energy efficiency \cite{2016Bjornson}. Nevertheless, the presence of spatial correlation among the multiple massive \gls{mimo} channels reduces the level of favorable propagation and channel hardening \cite{2019Bjornson,2020Sanguinetti}. 

Physical-layer multicasting is an efficient technique for group communication enabling multiple users, which demand for the same service and share the same time-frequency resources, to be simultaneously served by a single transmission. 
\Gls{lte-a} systems fully support broadcast/multicast transmissions through the use of the \gls{embms} \cite{Lecompte2012,2016delaFuente,2017Araniti}. The \gls{embms} is implemented as an \gls{lte-a} subsystem to share the physical resources between unicast and multicast transmissions. The standard allows the system an efficient resource utilization when multiple users simultaneously demand for the same content. Recently, \cite{2020Garro} proposes the \gls{5gnr} \gls{mm} to enhance the utilization of multicast in the \gls{5gnr} Release 17. \gls{5gnr} \gls{mm} provides a flexible, dynamic, and seamless switching between unicast and multicast or broadcast transmissions and traffic multiplexing under the same radio structures.

Lately, there is an increasing interest in multicast massive \gls{mimo} communications both in \gls{mmwave} and sub-6 GHz frequency bands. Combining multicasting and massive \gls{mimo} technology results in higher spectral efficiency, provided that an effective precoding scheme is implemented. This combination improves the content distribution in the demanding scenario of new wireless services and applications (e.g., video conference, mobile commerce, intelligent transportation systems, virtual and augmented reality). Additionally, coded multicasting is employed in the emerging coded caching techniques for content delivery of individual data requests to reduce wireless traffic. Thus, this area of application expands the potential of massive MIMO multicasting in the context of the delivery of content centric wireless networks \cite{2019Dong}. In this regard, there are many works in the literature proposing several fully digital precoding strategies to optimize both single- and multi-group multicast transmissions for sub-6 GHz massive \gls{mimo} systems. The authors in \cite{2013Yang} propose a multicast massive \gls{mimo} framework wherein a unique pilot sequence is assigned to all the multicast users receiving the same content. Accordingly, a joint power allocation strategy for downlink data transmission and uplink pilot transmission is then performed to maximize the equalized per-user throughput. Sadeghi \textit{et al.} extended the work in \cite{2013Yang} by considering joint multi-group multicast and unicast services in multi-cell massive \gls{mimo} deployments~\cite{2018SadeghiTWC1,2018SadeghiTWC2}, and by developing a low-complexity solution for multicasting\cite{2017SadeghiTWC}. In \cite{2019Dong}, the authors present a framework to achieve the optimal multicast beamforming. The low-dimensional structure in the optimal solution benefits the numerical computation in large antenna systems.

Recent research in multicast transmissions has also focused on analog and hybrid beamforming techniques for multicast massive \gls{mimo} systems operating at the \gls{mmwave} bands. The authors in \cite{2018SadeghiEW} present a hybrid precoding structure for multi-group physical layer multicasting. This strategy significantly reduces the number of required radio frequency (RF) chains to achieve the performance of any fully-digital precoder. In \cite{2019Biason}, the authors show how to shape the beams to deliver multicast information to the users in \gls{mmwave}. They demonstrate that restricting the wireless links to be unicast may only be suboptimal. Besides, the authors in \cite{2020Samuylov} develop a mathematical framework to estimate the parameters of the \gls{mmwave} \glspl{bs} for handling a mixture of multicast and unicast sessions. This framework allows the network designers to achieve a lower bound on the required density of the \glspl{bs}.

Existing works on multicast massive \gls{mimo} systems operating at the sub-6 GHz frequency bands, assume uncorrelated Rayleigh fading channel models, mainly for mathematical tractability convenience. However, spatial channel correlation always appears in practice, and its impact on the performance of massive \gls{mimo} systems is significant~\cite{Sanguinetti2020}. Especially, favorable propagation and channel hardening might be hindered by the spatial channel correlation.  
Motivated by the above considerations, in this work, we consider spatially correlated fading channels in a single-cell massive \gls{mimo} system underlaying multicast communications. Our novel technical contributions consist in:
\begin{itemize}
    \item A multicast user subgrouping strategy based on the large-scale spatial channel correlation characteristics. The subgroups are determined by the level of the users' channel mutual-orthogonality. Our subgrouping strategy reduces the pilot contamination among the users belonging to different multicast subgroups. As a result, the channel estimates employed for designing the precoding vectors are more accurate, and the precoding becomes more effective. This inevitably improve the \gls{se}.
    \item An optimal \gls{mmf} \gls{dl} power control to maximize the equalized \gls{se} among the different multicast subgroups.
    \item A heuristic and iterative \gls{mmf} power control strategy for intra-subgroup \gls{ul} pilot transmission, assuming correlated Rayleigh fading channels. Our iterative strategy aims to maximize the minimum \gls{se} of each multicast subgroup and builds on the closed-form expression for the max-min fairness uplink powers given in~\cite{2018SadeghiTWC1}, which was optimal only under the assumption of uncorrelated Rayleigh fading channels. 
    \item An exhaustive simulation campaign that quantifies the benefits provided by our subgrouping and power allocation strategies, under different system setups. 
    The proposed subgrouping-based multicasting is compared with the conventional multicasting, and unicast transmissions.
\end{itemize}

The remainder of this paper is organized as follows. In Section II, we present the system model, including a detailed description of the correlated Rayleigh fading channel model with its angular representation, and the \gls{ul} training phase. In this section, we also derive the expressions of the effective \gls{sinr} and \gls{se} for the downlink. In Section III, we present the proposed multicast user subgrouping model. Section IV describes the proposed optimal \gls{mmf} power allocation scheme for the downlink, and the heuristic power allocation scheme for the \gls{ul} pilots. In Section V, we show the results of our simulations to assess the technical soundness of the proposed strategies. Finally, Section VI concludes the paper by discussing  the importance of exploiting the spatial channel correlations in multicast massive \gls{mimo} systems and gives some clues on future research directions.

We use boldface lower (upper) case letters for vectors (matrices). Calligraphy, uppercase letters are used to denote sets, and $|\mathcal{A}|$ denotes the cardinality of set $\mathcal{A}$. The superscripts $(\cdot)\trans$, $(\cdot)^{\ast}$ and $(\cdot)\herm$ denote the transpose, the conjugate and the conjugate transpose (Hermitian) operators, respectively. $\mathbb{C}$ represents the set of complex numbers. $\mathbb{E}\{\cdot\}$ and $\mathbb{V}\{\cdot\}$ denote the expectation and the variance operators. $\text{tr}(\boldsymbol{A})$ denotes the trace of matrix $\boldsymbol{A}$.
A circularly symmetric complex Gaussian distribution with mean $\boldsymbol{\mu}$ and covariance matrix $\boldsymbol{\Sigma}$ is denoted by $\mathcal{CN}(\boldsymbol{\mu},\boldsymbol{\Sigma})$.
$\mathbf{I}_N$ represents the $N \times N$ identity matrix. $\norm{\boldsymbol{a}}_2$ denotes the $\ell_2$-norm of the vector $\boldsymbol{a}$.


\section{System Model}
\label{sec:system_model}
We consider a single multicast downlink transmission in a single-cell sub-6 GHz massive \gls{mimo} system, with fully digital precoding. This system consists of a macro \gls{bs} equipped with a \gls{ula} with $M$ transmit antennas, that delivers a multicast service to $K$ single-antenna users. The set of multicast users is denoted by $\mathcal{K}$, with $\mathcal{K} = \{1,\ldots, K\}$. 
The users are grouped into $G$ disjoint subgroups, and we denote the set of the indices of the users in subgroup $g$ by $\mathcal{K}_g$. We detail the proposed subgrouping strategy in Section \ref{sec:subgrouping}.

\subsection{Channel Model}
We assume a conventional block fading channel model, wherein the channel is time-invariant and frequency flat within a coherence block, and varies over different coherence blocks. 
Let $\boldsymbol{h}_{gk} \in \mathbb{C}^{M \times 1}$ be the channel response vector, in an arbitrary coherence block\footnote{For the sake of brevity, we omit the index identifying the coherence block.}, between the massive \gls{mimo} \gls{bs} and the single-antenna multicast user $k$ included in subgroup $g$. 
Under the assumption of spatially correlated Rayleigh fading channels, we have
\begin{equation}
    \begin{split}
        \boldsymbol{h}_{gk}  \sim \mathcal{CN}(0,\boldsymbol{R}_{gk}), \; \forall~g, k \in \mathcal{K}_g,
    \end{split}
\end{equation} 
where $\boldsymbol{R}_{gk} \in \mathbb{C}^{M \times M}$ is the positive semi-definite spatial covariance matrix of user $k$ in subgroup $g$, capturing the macroscopic propagation effects, namely the large-scale fading phenomena, including path-loss, shadow fading, and spatial correlations. While, the small-scale fading follows a complex Gaussian distribution.
Spatially correlated fading appears when either the channel gain and the channel direction are not uncorrelated, or the distribution of the channel directions is not uniformly over the unit-sphere in $\mathbb{C}^M$. The normalized trace of $\boldsymbol{R}_{gk}$ provides the large-scale fading coefficient ${\beta}_{gk}$, given by 
\begin{equation}
    \begin{split}
        {\beta}_{gk}  = \frac{1}{M} \text{tr}\left(\boldsymbol{R}_{gk}\right),
    \end{split}
    \label{eq:beta}
\end{equation}  
which gives a normalized measure of the channel gain.
The covariance matrix $\boldsymbol{R}_{gk}$ is characterized by the azimuth angles from the \gls{bs} to the users. 
The \gls{bs} receives from user $k$ a signal that consists of a superposition of $N$ multipath components. 
We reasonably assume the absence of scattering around the \gls{bs} (whose \gls{ula} antennas are placed at tens of meters on the ground level), and thereby all the multipath components originate from a scattering cluster around the user. 
Thus, each multipath component reaches the \gls{bs} as a planar wave from a specific angle of arrival (AoA) $\varphi_k(n) \in \left[\Phi_k,\Phi_k + \phi_k\right]$ for $n=1,\ldots,N$, where $\phi_k$ is a random deviation from the nominal angle $\Phi_k$. The standard deviation in radians characterizing the AoA is called \gls{asd}. Hence, the channel vector of user $k \in \mathcal{K}_g$ is given by
\begin{equation}
    \begin{split}
        \boldsymbol{h}_{gk} = \sum_{n=1}^N \rho_k(n)\boldsymbol{a}_k\left(\varphi_k(n)\right),
    \end{split}
\end{equation}  
where $\rho_k(n) \sim \mathcal{CN}(0,{\beta}_{gk}(n))$ represents the complex-valued gain of the $n$-th physical path, with strength ${\beta}_{gk}(n)$. The steering vector $\boldsymbol{a}_k\left(\varphi_k(n)\right) \in \mathbb{C}^{M}$ for an arbitrary AoA over the $n$-path, $\varphi_k(n)$, is given by
\begin{equation}
    \begin{split}
        \boldsymbol{a}_k\left(\varphi_k(n)\right) = \left[1 \ \ e^{j2\pi\delta\cos{\varphi_k(n)}} \ \ \ldots \ \ e^{j2\pi\delta(M-1)\cos{\varphi_k(n)}}\right]\trans,   
    \end{split}
\end{equation}  
where $\delta$ is the distance between adjacent antennas, normalized by the wavelength.
Both the nominal angle $\Phi_k$ and the \gls{asd} determine the spatial channel correlation. The \gls{bs} estimates the covariance matrix of each user on the large-scale fading time scale (i.e., over multiple coherence blocks)~\cite{Bjornson2016a,Neumann2018,Upadhya2018}. Therefore, we can reasonably assume that $\boldsymbol{R}_{gk}, \forall~g, k \in \mathcal{K}_g$, is known at the \gls{bs}.

\subsection{Multicast massive \gls{mimo} subgrouping}
\label{sec:subgrouping}
In this section, we detail the proposed multicast user subgrouping model, while the proposed criterion used to form the subgroups is described in Section~\ref{sec:mam}. 
As we assume that all the $K$ users in the system demand for the same service, they can be jointly and simultaneously served by a single multicast transmission, provided that the transmission rate is not higher than the rate supported by the user with the worst channel conditions. 
Clearly, this might severely penalize the performance of the users with good channel conditions, which could have rather experienced a higher quality-of-service (QoS) with a unicast transmission. In this regard, we propose to group the $K$ users into $G$ disjoint subgroups, each one served by a multicast transmission with a properly selected rate. 
Let $K_g=|\mathcal{K}_g|$ be the number of users in subgroup $g$, it holds that $K = \sum_{g=1}^G K_g.$
The purpose of using this user subgrouping strategy is to increase the sum rate of the multicast service. 


\subsection{Uplink Training}
massive \gls{mimo} systems conveniently operate in \gls{tdd} mode to limit the \gls{csi} acquisition overhead, which would otherwise scale also with the large number of \gls{bs} antennas~\cite{2016Marzetta}. We assume the \gls{bs} estimates the \gls{csi} from pilot sequences simultaneously sent by all the users during the \gls{ul} training stage. As adopting mutually-orthogonal pilots for all the users might be either expensive or not possible due to the finite length of the coherence block, we assume pilot reuse among the users. Specifically, let $\tau_{\mathrm{p}} < K$ be the pilot sequence length as well as the length of the uplink training stage,  then each user is assigned a pilot sequence randomly chosen among $\tau_{\mathrm{p}}$ available mutually-orthogonal pilots. In \cite{2013Yang,2017SadeghiTWC}, the authors assume that the users in the same multicast group are assigned the same pilot sequences. We consider a similar approach but at the multicast subgroup level, as we assume only one multicast group. Specifically, we propose that users in the same multicast subgroup share the same pilot sequence, while mutually-orthogonal pilots are assigned among different multicast subgroups.
Note that, as co-pilot users have linearly dependent channel estimates~\cite{2016Marzetta,2017Bjornsonbook}, the \gls{bs} cannot separate the users of the same subgroup in the spatial domain. 

Let $\boldsymbol{\Psi} = \left[\boldsymbol{\psi}_1,\ldots,\boldsymbol{\psi}_G\right] \in \mathbb{C}^{\tau_{\mathrm{p}} \times G}$ be the pilot matrix, where $\boldsymbol{\psi}_g$ is the pilot sequence assigned to each user $k \in \mathcal{K}_g$. Without loss of generality, we set $\tau_{\mathrm{p}} = G$ to obtain $G$ mutually orthogonal \gls{ul} pilot sequences, satisfying $\boldsymbol{\Psi}\trans\boldsymbol{\Psi}^{\ast}=\tau_p\mathbf{I}_G$.\footnote{$G$ is known at the uplink training stage.} Therefore, the \gls{ul} pilot signal received at the \gls{bs} is
\begin{gather}
    \boldsymbol{Y} = \sum^G_{g=1}\sum_{k \in \mathcal{K}_g} \sqrt{q_{gk}}\boldsymbol{h}_{gk}\boldsymbol{\psi}\trans_g + \boldsymbol{N},
\end{gather}
where $q_{gk}$ is the pilot transmit power of user $k \in \mathcal{K}_g$, and $\boldsymbol{N} \in \mathbb{C}^{M \times \tau_p}$ is \gls{awgn} with i.i.d. elements $\mathcal{CN}(0,\sigma^2)$. To estimate the channel of user $k$ in subgroup $g$, the \gls{bs} 
correlates the received \gls{ul} training signal with the corresponding pilot sequence $\boldsymbol{\psi}_g^{\ast}$, the \gls{bs} obtains the contaminated pilot signal of the user $k \in \mathcal{K}_g$ as
\begin{gather}
    \boldsymbol{y}_{gk}^{\mathrm{ul}} = \tau_{\mathrm{p}} \sqrt{q_{gk}}\boldsymbol{h}_{gk} + \tau_{\mathrm{p}} \sum_{j \in \mathcal{K}_g \setminus \{k\}} \sqrt{q_{gj}}\boldsymbol{h}_{gj} + \boldsymbol{n}_k,
\end{gather}
where $\boldsymbol{n}_k = \boldsymbol{N} \boldsymbol{\psi}^{\ast}_g  \sim \mathcal{CN}(0,\sigma^2\mathbf{I}_M)$ is the \gls{awgn} at the \gls{ul}. Provided that the users' covariance matrices are known a priori, the \gls{bs} can estimate the channel response vector $\boldsymbol{h}_{gk}$ using the \gls{mmse} estimation method \cite[Sec. 3.2]{2017Bjornsonbook}, as
\begin{gather}
    \hat{\boldsymbol{h}}_{gk} = \sqrt{q_{gk}}\boldsymbol{R}_{gk}\Bigg(\taup \sum_{j \in \mathcal{K}_g} q_{gj} \boldsymbol{R}_{gj} + \sigma^2\mathbf{I}_{M}\Bigg)^{\!\!\!-1}\boldsymbol{y}_{gk}\UL.
\label{eq:h_gk}
\end{gather}

The estimation error $\tilde{\boldsymbol{h}}_{gk} = \boldsymbol{h}_{gk} - \hat{\boldsymbol{h}}_{gk}$ has correlation matrix $\tilde{\boldsymbol{R}}_{gk} = \mathbb{E} \{\tilde{\boldsymbol{h}}_{gk}(\tilde{\boldsymbol{h}}_{gk})\herm\}$, given by
\begin{align}
    \tilde{\boldsymbol{R}}_{gk} \!=\! \boldsymbol{R}_{gk} \!-\! q_{gk}\tau_p\boldsymbol{R}_{gk}\Bigg(\!\taup\! \sum_{j \in \mathcal{K}_g} q_{gj} \boldsymbol{R}_{gj} \!+\! \sigma^2\mathbf{I}_{M}\!\Bigg)^{\!\!\!-1}\!\boldsymbol{R}_{gk}.
\label{eq:R_est_error}
\end{align}

Let $\boldsymbol{h}_g$ denote the composite channel of the multicast subgroup $g$, which consists of a linear combination of the channels of all the users in $\mathcal{K}_g$, given by 
\begin{gather}
\boldsymbol{h}_g = \taup\displaystyle\sum_{k \in \mathcal{K}_g}\sqrt{ q_{gk}}\boldsymbol{h}_{gk}.
\end{gather}
Then, the composite \gls{mmse} channel estimate is given by
\begin{gather}
    \hat{\boldsymbol{h}}_{g} = \taup\sum_{k \in \mathcal{K}_g} q_{gk} \boldsymbol{R}_{gk}\Bigg(\!\taup\! \sum_{j \in \mathcal{K}_g} q_{gj} \boldsymbol{R}_{gj} \!+\! \sigma^2\boldsymbol{I}_{M}\!\Bigg)^{\!\!\!-1}\!\boldsymbol{y}_{gk}\UL.
\label{eq:h_g}
\end{gather}
The composite channel estimate is then used to design the corresponding precoding vector for all the users in subgroup $g$ capitalizing on the channel reciprocity resulting from the TDD operation mode. 



\subsection{Downlink Data Transmission and Spectral Efficiency}

The \gls{bs} performs $G$ multicast transmissions, that is a multicast transmission per subgroup, by employing for each one a unique combination of precoding vector and coding scheme. For instance, let $\boldsymbol{w}_g$ denote the precoding vector intended for the multicast subgroup $g$, in case of \gls{mr} transmission strategy, we have
\begin{equation}
    \begin{split}
\boldsymbol{w}_g = \frac{\hat{\boldsymbol{h}}_{g}}{\norm{\hat{\boldsymbol{h}}_{g}}_2},    \end{split}
    \label{eq:MR}
\end{equation}
where $\boldsymbol{w}_g \in \mathbb{C}^{M \times 1}$ and  $\mathbb{E}\left[\norm{\boldsymbol{w}_g}^2\right]=1$.
Let $x_g$ be the data symbol intended for user $k \in \mathcal{K}_g$, with $\norm{x_g}^2_2 = 1$, and $\mathbb{E}\{x_g x_c^{\ast}\}=0$, $\forall~g \neq c$. In addition, $p_g$ denotes the transmit power of the symbol $x_g$. Then, assuming that the users does not have knowledge of the instantaneous effective downlink channel, i.e., $\boldsymbol{h}\herm_{gk}\boldsymbol{w}_g$, but rather they have access only to the channel statistics, i.e., $\mathbb{E}\left\{\boldsymbol{h}\herm_{gk}\boldsymbol{w}_g\right\}$, we can express the \gls{dl} data signal received at user $k \in \mathcal{K}_g$ as
\begin{align}
        y_{gk} &\!=\! \sqrt{p_g} \ \mathbb{E}\left\{\boldsymbol{h}\herm_{gk}\boldsymbol{w}_g\right\} x_g \!+\! \sqrt{p_g}\left(\boldsymbol{h}\herm_{gk}\boldsymbol{w}_g \! - \! \mathbb{E}\left\{\boldsymbol{h}\herm_{gk}\boldsymbol{w}_g\right\}\right) x_g \nonumber \\
        & \qquad \!+\! \sum^G_{\substack{c = 1 \\ c \neq g}} \sqrt{p_c} \ \boldsymbol{h}\herm_{gk}\boldsymbol{w}_c x_c + n_k,
    \label{eq:y_gk}
\end{align}    
where the first term denotes the desired signal, the second term is interference due to the user's lack of \gls{csi}, the third term denotes the inter-subgroup interference, and $n_k \sim \mathcal{CN}(0,\sigma_k^2)$ is the \gls{awgn} at the user $k$. Note that all the terms, with the exception of the first one, are random variables whose realizations are not known at the user. 

By invoking the capacity-bounding technique in \cite[Sec. 2.3.4]{2016Marzetta}, which treats the second, third and fourth term of \eqref{eq:y_gk} as effective uncorrelated noise\footnote{Recall that the data symbols intended for different subgroups are uncorrelated using for each one a unique combination of precoding  vector and coding scheme.}, a \gls{dl} achievable \gls{se} is given~by 
\begin{equation}
    \begin{split}
           \text{SE}_{gk} = \left(1 - \frac{\taup}{\tau}\right) \text{log}_2\left(1 + \gamma_{gk}\right),
     \end{split}
    \label{eq:SE_gk}
\end{equation}
where $\tau$ is the length of the coherence block, and $\gamma_{gk}$ is the effective \gls{sinr} given by
\begin{equation}
        \!\gamma_{gk} \!=\! \frac{p_g\Big|{\mathbb{E}\left\{\boldsymbol{h}\herm_{gk}\boldsymbol{w}_g\right\}}\Big|^2}{\displaystyle\sum_{c=1}^G p_c~\mathbb{E}\left\{\Big|{\boldsymbol{h}\herm_{gk}\boldsymbol{w}_c}\Big|^2\right\} \!-\! p_g\Big|{\mathbb{E}\left\{\boldsymbol{h}\herm_{gk}\boldsymbol{w}_g\right\}}\Big|^2 \!+\! \sigma_k^2},
    \label{eq:SINR_gk}
\end{equation}
where the expectations are with respect to the channel realizations, and the channel estimates inside the precoding vectors.
The expression in~\eqref{eq:SE_gk} describes an achievable \gls{se} for user $k$ in subgroup $g$, for any precoding strategy. Notice that closed-form expressions for the effective SINR can be easily derived by using tools from matrix theory when \gls{mr} and \gls{zf} precoding schemes are used. We omit these expressions as they are well known in the literature~\cite{2017Bjornsonbook}. Nevertheless, we stress that a subgroup is served by a single multicast transmission, which determines a shared \gls{dl} SE for all the users of the subgroup. In order for the data to be reliably decoded by all the users of the subgroup, the shared transmit SE must support the DL \gls{se} of the user with the worst channel conditions in the subgroup, namely the minimum \gls{se} among the users of the same subgroup:
\begin{equation}
    \begin{split}
           \text{SE}_{g} = \underset{k \in \mathcal{K}_g}{\text{min}} \text{SE}_{gk}, \quad \forall \, g.
     \end{split}
    \label{eq:SE_g}
\end{equation}
The expression in~\eqref{eq:SE_g} describes the effective DL \gls{se} for each user in subgroup $g$.

\section{MaMIMO multicasting with User Subgrouping}
\label{sec:mam}
A massive \gls{mimo} \gls{bs} can simultaneously deliver content to multiple multicast groups in the same time-frequency resources, by multiplexing those in the spatial domain. The literature of massive \gls{mimo} multicasting \cite{2013Yang,2017SadeguiTVT,2017SadeghiTWC,2018SadeghiTWC1,2019Dong} essentially presents two fundamental delivery strategies. 
The first option consists in serving each multicast user by a unicast transmission as in conventional massive \gls{mimo}. 
This approach leads to a significant increase of the frequency of the pilot reuse, hence of the level of pilot contamination in the system. 
Furthermore, high correlated channels lead to the necessity of many antennas at the \gls{bs} side to achieve the required spatial resolution. 

The second option consists in serving all the users demanding for the same service (which form a multicast group) by a single multicast transmission. In this case, the same pilot sequence is assigned to all the users of the multicast group. Hence, the \gls{bs} employs a single (composite) precoding vector for the multicast transmission. 
On one hand, this strategy significantly reduces the uplink training length as the requirements in term of mutually-orthogonal pilots are lower. This is particularly helpful in scenarios where the coherence blocks are relatively short, or as the number of users in the system grows. 
On the other hand, a multicast transmission for all the users may lead to a performance degradation as a common single transmission cannot fully achieve the desired SE of each single user, but it is rather constrained by the multicast user with the worst channel conditions. 

Hence, we propose an alternative strategy to deliver a multicast service in a massive \gls{mimo} system. This strategy consists, as already mentioned, in creating disjoint subgroups of multicast users. 
This approach has already shown high-quality results in traditional \gls{siso} and \gls{mimo} systems, using both wideband and subband channel information \cite{2013Araniti,2018delaFuente}. 
However, with respect to prior works, our contribution consists of a novel criterion adopted to form the subgroups. 
As a measure of the similarities among the user's channel characteristics, we consider the level of channel orthogonality between two different users in two different subgroups, quantified by the inner product of the normalized channels given by
\begin{equation}
    \begin{split}
    \frac{\boldsymbol{h}_{gk}\herm \boldsymbol{h}_{cj}}{\sqrt{\mathbb{E}\{\norm{\boldsymbol{h}_{gk}}^2\}\mathbb{E}\{\norm{\boldsymbol{h}_{cj}}^2\}}},
    \end{split}
    \label{eq:favorable_propagation}
\end{equation}
with $c \neq g$, and $k \neq j$. The smaller this value is, the greater the orthogonality between the two channel vectors is. This also represents a measure of the \emph{favorable propagation} between the channel vectors $\boldsymbol{h}_{gk}$, $\boldsymbol{h}_{cj}$ of any pair of multicast users $k \in \mathcal{K}_g$, $j \in \mathcal{K}_{c}$. If these two channel vectors are nearly orthogonal, then the corresponding inter-user interference is negligible. As the channel realizations $\boldsymbol{h}_{gk}$, $\boldsymbol{h}_{cj}$ are not known at the base station, such a metric cannot be computed. Moreover, even considering the metric~\eqref{eq:favorable_propagation} but replacing the true channel responses with the corresponding channel estimates is not practical, as the estimates vary every coherence block, and thereby the BS should re-configure the subgroups every coherence block (i.e., every few milliseconds). 
Hence, we rather take into account a related measure that varies on the large-scale-fading time scale, given by
\begin{align}
        &\mathbb{V}\left\{\frac{\boldsymbol{h}_{gk}\herm \boldsymbol{h}_{cj}}{\sqrt{\mathbb{E}\{\norm{\boldsymbol{h}_{gk}}^2\}\mathbb{E}\{\norm{\boldsymbol{h}_{cj}}^2\}}}\right\} \nonumber \\
        & \quad= \frac{\mathbb{V}\left\{\boldsymbol{h}_{gk}\herm \boldsymbol{h}_{cj}\right\}}{\mathbb{E}\{\norm{\boldsymbol{h}_{gk}}^2\}\mathbb{E}\{\norm{\boldsymbol{h}_{cj}}^2\}} =\frac{\text{tr}\left(\boldsymbol{R}_{gk} \boldsymbol{R}_{cj}\right)}{\text{tr}\left(\boldsymbol{R}_{gk}\right)\text{tr}\left(\boldsymbol{R}_{cj}\right)},
    \label{eq:cov_matrix}
\end{align}
which can be easily computed as the channel covariance matrices are known at the BS by assumption.
The proposed subgrouping criterion thus relies on a distance metric between two elements $\boldsymbol{R}_{gk}, \boldsymbol{R}_{cj} \in \mathbb{C}^{M \times M}$, that is on the similarities among the users' spatial correlation matrices. 
The BS groups users whose channels present a low degree of mutual-orthogonality, while users in different subgroups are characterized by high levels of mutual-orthogonality. 

The $K$-means algorithm and its multiple variants provide a simple method to efficiently divide the multicast users into disjoint subgroups \cite{2010Jain}. This algorithm aims at finding a partition of the $K$ users into $G$ subgroups \cite{2018Riera}, that minimizes the \gls{mse} between the empirical mean of all the users.
Deep learning strategies can be also used to achieve the optimal value of the number of subgroups, $G$, in a single-cell multicast service~\cite{Chukhno2021}. However, the calculation of the optimal number of multicast subgroups is out of the scope of this work.

\begin{figure}[!t]
\centering
\includegraphics[width=.7\linewidth]{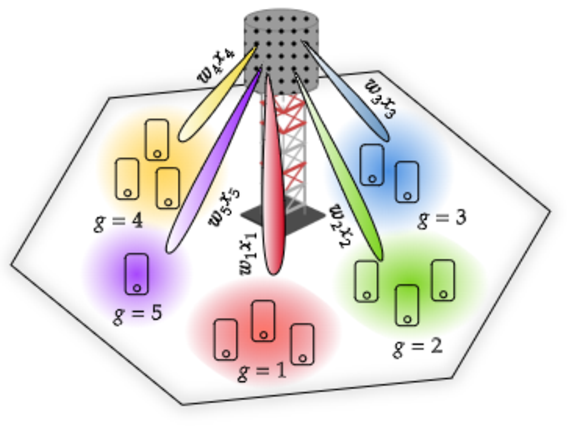}
\caption{Subgroup multicast massive MIMO transmissions with user subgrouping based on the users' channel spatial similarities.}
\label{fig:subgroupingscenario}
\end{figure}

Figure \ref{fig:subgroupingscenario} illustrates an instance of a multicast service in a single-cell massive \gls{mimo} system. The users are grouped based on the level of orthogonality of their spatial correlation matrices, namely users with similar spatial characteristics belong to the same subgroup. Multicast users of the same subgroup suffer of mutual interference due to pilot contamination, as the BS utilizes a common precoder towards them. The a priori knowledge at the BS of $\boldsymbol{R}_{gk}$ is used to allocate the multicast users into disjoint subgroups.

\section{Power Allocation}

Unlike in traditional \gls{siso} and \gls{mimo} systems, by leveraging the \textit{law of the large numbers}, closed-form expressions for an achievable \gls{se} can be derived, and thanks to the channel hardening phenomenon, these expressions become more reliable as the number of BS antennas grows. Hence, the theoretical analysis of massive \gls{mimo} is able to accurately predict the performance of practical systems. Besides, the derived closed-form expressions for the \gls{se} can be exploited to simplify optimal power control strategies, which can exclusively rely on the large-scale fading quantities.

\Gls{mmf} power control~\cite{2016Marzetta,2017Bjornsonbook} is unquestionably of practical interest in massive \gls{mimo} multicasting since it is an egalitarian power control strategy that equalizes the \glspl{se} of the users, and maximizes such equalized \gls{se}. Additionally, max product \gls{sinr} power allocation also provides a fairness strategy among the multicast users that results in a good trade-off among the achieved sum \gls{se} and the guaranteed \gls{se} delivered for the majority of the users \cite{2017Bjornsonbook}.

In our proposed subgrouping framework, we employ an uplink/downlink double MMF power allocation strategy. First, we propose a heuristic algorithm to achieve a nearly optimal \gls{ul} pilot power allocation that aims at maximizing the minimum \gls{se} in each single subgroup, for a given downlink power allocation. We dub this approach as \textit{intra-subgroup \gls{mmf}} power control. Then, an \textit{inter-subgroup} \gls{mmf} power control is performed to optimize the downlink power allocation, and maximizing the minimum \gls{se} of each single subgroup. We dub this approach as \textit{inter-subgroup  \gls{mmf}} power control.

\subsection{Intra-subgroup \gls{mmf} power control}

Optimizing the transmit powers of the \gls{ul} pilots leads to an improvement in the subgroup precoder design that provides a better equalization of the per-user SE throughout the same subgroup and, in turn, of the effective subgroup SE. 
Prior works~\cite{2013Yang,2018SadeghiTWC1,2018SadeghiTWC2} proposed a \gls{mmf} power control for the \gls{ul} pilot transmission, to maximize an equalized SE among all the multicast users. Specifically,~\cite{2018SadeghiTWC1} formulates a joint \gls{mmf} power control for downlink data transmission and uplink pilot transmission, and, capitalizing on the analytical tractability of the optimization problem under the assumption of uncorrelated Rayleigh fading channels, gives the optimal solutions for both the DL data and UL pilot transmit powers in closed form.

In this work, we propose a heuristic algorithm to achieve a nearly optimal \gls{ul} pilot power allocation that aims to maximize the minimum \gls{se} in each subgroup. For each multicast subgroup, this algorithm iteratively updates the UL pilot power control coefficients, $\{q_{gk}\}$, to increase the minimum per-user \gls{sinr} in each subgroup. Algorithm \ref{alg:Intrasubgroup-MMF} describes in detail the steps of the proposed heuristic intra-subgroup \gls{mmf} power control. This algorithm overlays an existing subgrouping configuration and runs for an arbitrary subgroup $g$. Its inputs are the users number in the subgroup $\mathcal{K}_g$, the subgroups number $G$, the channel covariance matrices $\{\boldsymbol{R}_{gk}\}$ and the channel error correlation matrices $\{\tilde{\boldsymbol{R}}_{gk}\}$ of all the multicast users, the DL and UL power budget $P_{dl}$ and $Q_{ul}$, and the pilot sequence length $\tau_p$. 

We run this algorithm prior to DL power allocation to calculate a close-to-optimal distribution of MMF UL pilot power in every subgroup. Note that the UL pilot power allocation and the DL power allocation are coupled problems. To circumvent this coupling, we employ in the UL pilot power control algorithm an a-priori DL power distribution based on the available large-scale-fading channel information as  
\begin{equation}
p_g = P_{dl}\frac{\beta_g^{\nu}}{\sum_{c=1}^G \beta_c^{\nu}},
\label{eq:p_g}
\end{equation}
where 
\begin{equation}
\beta_g = \frac{1}{M}\sum_{k \in \mathcal{K}_g} \Big[\text{tr}\left(\boldsymbol{R}_{gk}\right)-\text{tr}\left(\tilde{\boldsymbol{R}}_{gk}\right)\Big],
\end{equation} 
$P_{\mathrm{dl}}$ is the \gls{dl} power budget at the BS per coherence block, and the parameter $\nu \in [-1,1]$ establishes the power allocation fairness. Such a power control strategy is also known as \textit{fractional power control}, and it is used in the LTE-A standard, for instance.

For initializing the \gls{ul} pilot power control coefficients we choose a sub-optimal\footnote{Note that, this choice is optimal in case of uncorrelated Rayleigh fading channel responses.} power allocation given by~\cite{2018SadeghiTWC1}, according to which 
\begin{equation}
q_{gk} = \frac{1+\beta_{gk}p_g}{\beta_{gk}^2} \Upsilon_g, \quad \forall \, g, k \in \mathcal{K}_g,
\label{eq:q_uncorr}
\end{equation}
where $\Upsilon_g = \underset{k \in \mathcal{K}_g}{\min}\dfrac{Q_{\mathrm{ul}}\beta_{gk}^2}{1+\beta_{gk}p_g}$,
and $Q_{\mathrm{ul}}$ being the UL transmit power budget for each user. 

This initial set of transmit powers for the UL pilot transmission determines the channel estimates $\{\hat{\boldsymbol{h}}_{gk}\}$ via eq.~\eqref{eq:h_gk}, the composite channel estimates $\{\hat{\boldsymbol{h}}_{g}\}$ via eq.~\eqref{eq:h_g}, the precoding vectors $\{\boldsymbol{w}_g\}$ via eq.~\eqref{eq:MR}, and lastly the effective DL SINR per user via eq.~\eqref{eq:SINR_gk} which is based on the large-scale channel statistics. 
In Algorithm \ref{alg:Intrasubgroup-MMF}, the \emph{while} loop performs an iterative search of the suboptimal $q_{gk}^{\star}$ that equalizes the \gls{sinr}s of all the users in subgroup $g$. The loop starts by updating the values of $q_{gk}$ for each user according to the iteration rule $$q_{gk} \leftarrow \mu_{gk}\frac{q_{gk}^{\star}}{\max\limits_{k \in \mathcal{K}_g}q_{gk}^{\star}},$$
where $\mu_{gk}$ represents the \textit{step size}. Then, the set consisting of the channel estimates, the composite channel estimates, the precoding vectors, the effective DL SINR per user are updated accordingly. %
\begin{figure}[!t]
\centering
\includegraphics[width=\linewidth]{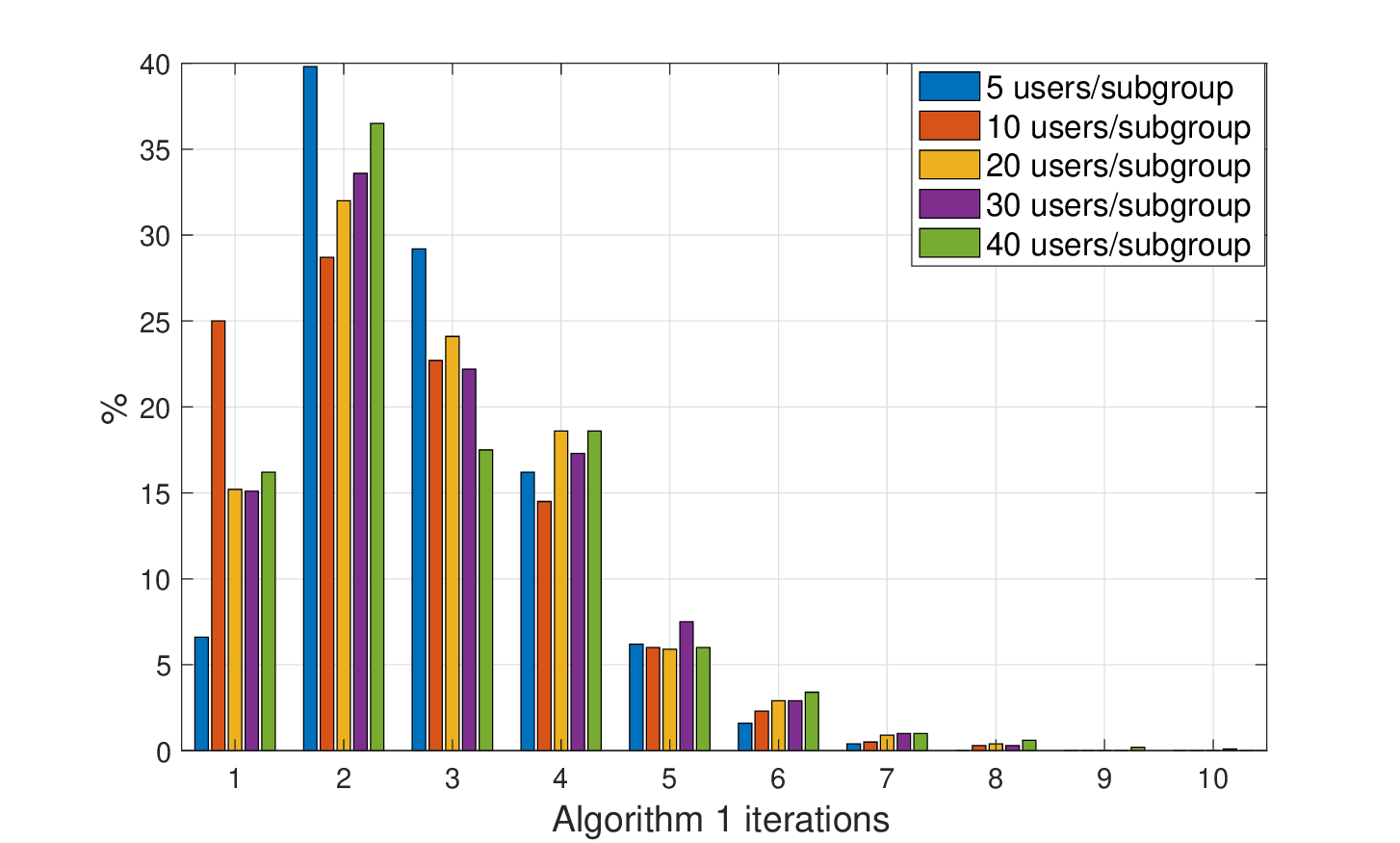}
\caption{Number of iterations needed for Algorithm 1 to converge for different numbers of users per subgroup. One iteration indicates that the output of Algorithm 1 is exactly given by Eq. (20). These results are obtained for 500 network snapshots with $M=64$, $P_{\mathrm{dl}}=33$ dBm, $Q_{\mathrm{ul}}=20$ dBm, $1$ cluster of $5$ m radius.}
\label{fig:iterations}
\end{figure}

If the new value of the minimum \gls{sinr} achieved in the subgroup $g$ is larger than the one at the previous iteration, stored into $\Gamma_{\mathrm{min}}$, then the set of coefficients $\{q_{gk}\}$ is a sub-optimal solution candidate, and we update $\Gamma_{\mathrm{min}}$ and the suboptimal UL pilot power control coefficients, $q_{gk}^{\star}$. Else, we finish the iterative search achieving a close-to-optimal MMF UL pilot power control. This greedy algorithm converges in very few iterations, as is illustrated in Figure \ref{fig:iterations}, where we observe that Algorithm 1 converges in less than 5 iterations with high probability, regardless of the number of users per subgroup. 
\begin{algorithm}[!t]
\caption{Iterative greedy algorithm for the intra-subgroup \gls{mmf} power control}\label{alg:Intrasubgroup-MMF}
\begin{algorithmic}
  \small	
    \State {\underline{\textbf{Input}:} \ $\taup$, $P_{\mathrm{dl}}$, $Q_{\mathrm{ul}}$, $\mathcal{K}_g$, $\{\boldsymbol{R}_{gk}\}, \{\tilde{\boldsymbol{R}}_{gk}\} \ \forall k \in \mathcal{K}$}
    \State {\underline{\textbf{Initialization}}}
    \State {$\vartheta \leftarrow 1$}
    \State {Calculate $p_g$ via (\ref{eq:p_g});}
    \State {Calculate $q_{gk}^{\star} \ \ \forall k \in \mathcal{K}_g$ via (\ref{eq:q_uncorr})}
    \State {Calculate $\hat{\boldsymbol{h}}_{gk}(q_{gk}^{\star}) \ \ \forall k \in \mathcal{K}_g$ via (\ref{eq:h_gk});}
    \State{Calculate $\hat{\boldsymbol{h}}_{g}(q_{gk}^{\star})$ via (\ref{eq:h_g});}
    \State{Calculate $\boldsymbol{w}_{g}(q_{gk}^{\star})$ via (\ref{eq:MR});}
    \State{Calculate $\gamma_{gk}^{\star}(q_{gk}^{\star}) \ \ \forall k \in \mathcal{K}_g$ via (\ref{eq:SINR_gk})};
    \State $\mu_{gk} \leftarrow \frac{\min\limits_{k \in \mathcal{K}_g}\gamma_{gk}^{\star}(q_{gk}^{\star})}{\gamma_{gk}^{\star}(q_{gk}^{\star})}, \, \forall k \in \mathcal{K}_g$
    \State $\Gamma_{\mathrm{min}} \leftarrow \min\limits_{k \in \mathcal{K}_g} \gamma_{gk}^{\star}(q_{gk}^{\star})$
    \While{$\vartheta$}
        \State $q_{gk} \leftarrow \mu_{gk}\frac{q_{gk}^{\star}}{\max\limits_{k \in \mathcal{K}_g} q_{gk}^{\star}}, \, \forall k \in \mathcal{K}_g$
        \State Calculate $\hat{\boldsymbol{h}}_{gk}(q_{gk}) \ \ \forall k \in \mathcal{K}_g$ via (\ref{eq:h_gk}); 
        \State Calculate $\hat{\boldsymbol{h}}_{g}(q_{gk})$ via (\ref{eq:h_g});
        \State Calculate $\boldsymbol{w}_{g}(q_{gk})$ via (\ref{eq:MR});
        \State Calculate $\gamma_{gk}(q_{gk}) \ \ \forall k \in \mathcal{K}_g$ via (\ref{eq:SINR_gk});
        \If{$\min\limits_{k \in \mathcal{K}_g}\gamma_{gk}(q_{gk}) > \Gamma_{\mathrm{min}}$}
            \State $\Gamma_{\mathrm{min}} \leftarrow \min\limits_{k \in \mathcal{K}_g}\gamma_{gk}(q_{gk})$
            \State $q_{gk}^{\star} \leftarrow q_{gk}, \, \forall k \in \mathcal{K}_g$ 
        \Else
            \State $\vartheta \leftarrow 0$
        \EndIf
    \EndWhile
    \State \underline{\textbf{Output}:}  \ \{$q_{gk}^{\star}$\} 
\end{algorithmic}
\end{algorithm}

\subsection{Inter-subgroup \gls{mmf} power control}

Let us consider that the intra-subgroup \gls{mmf} power control equalizes the \gls{sinr} of the users belonging to the same subgroup. Consequently, we define the \gls{mmf} optimization problem among multicast subgroups subject to the \gls{dl} power constraints as
\begin{equation}
    \begin{split}
        \mathcal{P}_1: \hspace{0.4cm} &\underset{\{p_g \geq 0 \}}{\text{maximize}}\hspace{0.2cm} \underset{g, \, k \in \mathcal{K}_g}{\min} \hspace{0.3cm} \text{SE}_{gk}\\
         & \hspace{0.5cm} \text{s.t.} \hspace{0.9cm} \sum\nolimits^G_{g=1} p_{g} \leq P_{\mathrm{dl}}.\\
   		\end{split}
   		\label{eq:MMF_problem}
\end{equation}
Eq.~\eqref{eq:SE_gk} can be rewritten as
\begin{equation}
    \begin{split}
        \text{SE}_{gk} = \left(1 - \frac{\taup}{\tau}\right) \log_2\left(1 +\frac{p_g \, a_{gk}}{\sum\nolimits^G_{c = 1} p_c \, b_{gkc} + \sigma_k^2}\right),
     \end{split}
    \label{eq:SE_gk2}
\end{equation}
where $a_{gk} \!=\! \Big|{\mathbb{E}\left\{\boldsymbol{h}\herm_{gk}\boldsymbol{w}_g\right\}}\Big|^2$, $b_{gkc} \!=\! \mathbb{E}\left\{\Big|\boldsymbol{h}\herm_{gk}\boldsymbol{w}_{c}\Big|^2\right\},~c \neq g$, and $b_{gkg} \!=\! \mathbb{E}\left\{\Big|\boldsymbol{h}\herm_{gk}\boldsymbol{w}_{g}\Big|^2\right\} - a_{gk}$.
As the logarithmic function is monotone increasing, maximizing the minimum \gls{se} is equivalent to maximizing the minimum \gls{sinr}. Therefore, we can rewrite $\mathcal{P}_1$ in epigraph form as
\begin{equation}
    \begin{split}
        \mathcal{P}_2: \hspace{0.2cm} &\underset{\{p_g \geq 0 \}}{\text{maximize}}\hspace{0.3cm} \Gamma\\
         & \hspace{0.4cm} \text{s.t.} \hspace{0.7cm} \frac{p_g a_{gk}}{\sum\nolimits^G_{c = 1} p_c b_{gkc} + \sigma_k^2} \geq \Gamma \hspace{0.3cm} \forall g, k \in \mathcal{K}_g \\
         &\hspace{1.5cm} \sum\nolimits^G_{g = 1} p_g \leq P_{\mathrm{dl}},\\
   		\end{split}
   		\label{eq:MMF_problem2}
\end{equation}
where $\Gamma$ is an auxiliary variable that serves as SINR target. 
$\mathcal{P}_2$ is non-convex since the \gls{sinr} constraint is neither convex nor concave with respect to the transmit powers $\{p_g\}$. To overcome such non-convexity, we use a \gls{sca}. Note that for a fixed value of $\Gamma \geq 0$, the \gls{sinr} constraint in $\mathcal{P}_2$ becomes linear \[ p_g a_{gk} \geq \Gamma\left(\displaystyle\sum_{c=1}^G p_{c} b_{gkc} + \sigma_k^2\right).\]

\begin{algorithm}[!t]
\caption{SCA algorithm for optimal \gls{mmf} power control in massive \gls{mimo} multicasting with user subgrouping}\label{alg:SCA-MMF}
\begin{algorithmic}
  \small	
    \State {\underline{\textbf{Input}:} \ \ $\{a_{gk}\}$, \ $\{b_{gkc}\}$, $P_{\mathrm{dl}}$}
    \State {\underline{\textbf{Initialization}:}}
    \State {$\Gamma_{\mathrm{min}} \leftarrow 0$}
    \State {$\Gamma_{\mathrm{max}} \leftarrow \underset{g,k}{\min}\left(\frac{P_{\mathrm{dl}} \, a_{gk}}{\sigma_k^2}\right)$}
    \State {$p^{\star}_g \leftarrow P_{\mathrm{dl}}/G, \ \forall g$}
    \While{$\Gamma_{\mathrm{max}} - \Gamma_{\mathrm{min}} > \varepsilon$} 
        \State {$\Gamma = (\Gamma_{\mathrm{max}} + \Gamma_{\mathrm{min}})/2$}
        \If {(\ref{eq:MMF_problem2}) is feasible}
            \State {$\Gamma_{\mathrm{min}} \leftarrow \Gamma$}
            \State {$p_g^{\star} \leftarrow p_g, \ \forall g $}
        \Else
            \State {$\Gamma_{\mathrm{max}} \leftarrow \Gamma$}
        \EndIf
    \EndWhile
    \State {\underline{\textbf{Output}:} \ $\{p_g^{\star}\}$}
\end{algorithmic}
\end{algorithm}
Hence, if $\Gamma$ is fixed, $\mathcal{P}_2$ becomes a linear feasibility program, and the optimal solution can be efficiently computed by using interior-point methods (e.g., the CVX toolbox~\cite{2014cvx}).
Letting $\Gamma$ vary over an \gls{sinr} search range $\{\Gamma_{\mathrm{min}}, \Gamma_{\mathrm{max}}\}$, the optimal solution can be efficiently computed by using the bisection method \cite{2004Boyd}, in each step solving the corresponding linear feasibility problem for a fixed value of $\Gamma$.
Algorithm \ref{alg:SCA-MMF} describes the \gls{sca} algorithm providing the power allocation that maximizes the minimum \gls{sinr} among the multicast subgroups. We dub this approach as \textit{inter-subgroup \gls{mmf}} power control. 
In algorithm \ref{alg:SCA-MMF}, the coefficients $\{a_{gk}\}$ and $\{b_{gkc}\}$, and $P_{\mathrm{dl}}$ are the inputs. Firstly, we initialize the range of $\Gamma$ values. The \emph{while} loop performs a bisection search over the values of $\Gamma \in \{\Gamma_{\mathrm{min}}, \Gamma_{\mathrm{max}}\}$, whose range limits are updated in each iteration according to the result from the feasibility linear program (\ref{eq:MMF_problem2}). Specifically, if (\ref{eq:MMF_problem2}) is feasible, $\Gamma_{\mathrm{min}}$ takes the value of the $\Gamma$ candidate, else $\Gamma_{\mathrm{max}}$ does. The \emph{while} loop ends when the difference between $\Gamma_{\mathrm{max}}$ and $\Gamma_{\mathrm{min}}$ is smaller than an accuracy threshold $\epsilon$, which is a parameter design (generally in the order of $10^{-6}$). Algorithm \ref{alg:SCA-MMF} finally returns the optimal power allocation given by the last feasible set of the power coefficients $\{p^{\star}_g\}$.


\section{Numerical results}
\label{sec:results}

In this section, we assess the performance of our multicast massive \gls{mimo} subgrouping strategy, the \emph{inter-subgroup} and the \emph{intra-subgroup \gls{mmf}} power control schemes. In our simulations, we deploy a single cell served by a macro \gls{bs}, which is equipped with a \gls{ula}. Within the nominal coverage area, we drop the users in clusters uniformly at random and compute the nominal angles from each of them to the \gls{bs}. Users' locations and nominal angles with respect to the ULA determine the spatial correlations between the channel vectors. This deployment represents a snapshot of the network. We run the proposed UL pilot/DL power control schemes and collect the spectral efficiency results for each snapshot. These results are averaged over $500$ network snapshots, i.e., random realizations of the large-scale fading coefficients, which guarantees that the CDFs do not deviate from their true value by more than $0.3\%$, with $95\%$ confidence. Table~\ref{tab:sim_parameters} shows a summary of the default simulation parameters.
\begin{table}[!t]
\renewcommand{\arraystretch}{1.2}
\centering
\caption{Simulation setup}
\begin{tabular}{l|c}
\bfseries Parameters & \bfseries Value\\
\hline
 {Number of BS antennas} &  {$64$}\\
 {Maximum DL transmit power} &  {$33$ dBm}\\
 {Maximum UL pilot transmit power} &  {$20$ dBm}\\
 {Path loss model [dB]} &  {$32.4 + 20\text{log}_{10}(f)$}\\
&  {$+ 37.6\log_{10}(d)$}\\
 {Angular standard deviation} &  {$10^{\circ}$}\\
 {User noise figure} &  {$7$ dB}\\
 {Noise power spectral density} &  {$-174$ dBm/Hz}\\
 {Cell radius} &  {$200$ m}\\
 {Cluster radius} &  {$5$ m}\\
 {Shadowing standard deviation} &  {$6$ dB}\\
 {Shadowing intra-cluster correlation} &  {$1$}\\
 {Shadowing inter-cluster correlation} &  {$0.1$}\\
 {Carrier frequency} &  {$2$ GHz}\\
 {Operating bandwidth} &  {$20$ MHz}\\
 {Channel coherence samples} &  {$200$}\\
 {Precoding strategy} & MR transmission\\
\hline
\end{tabular}
\label{tab:sim_parameters}
\end{table}
Our performance metric of interest is the \gls{sse} given by the sum of the \glspl{se} of the multicast subgroups as
    \begin{equation}
        \begin{split}
           \text{SumSE} = \sum\limits^G_{g = 1} \text{SE}_g = \sum\limits^G_{g = 1} K_g \underset{k \in \mathcal{K}_g}{\text{min}} \text{SE}_{gk}.
        \end{split}
        \label{eq:SumSE}
    \end{equation}
As we are using UL MMF pilot power control and DL MMF data power control, our strategy equalizes the SE of all the multicast users. As a result, the \gls{sse} can be calculated as  
\begin{equation}
        \begin{split}
           \text{SumSE} = K \times \overline{\text{SE}},
	\end{split}
        \label{eq:SumSE2}
    \end{equation}	
where $\overline{\text{SE}}$ is the per-user spectral efficiency $\forall k \in \mathcal{K}$.

\subsection{Benchmarking of the proposed \gls{mmf} power control algorithm and subgrouping strategy}

\begin{figure}[t]
\centering
\includegraphics[width=\columnwidth]{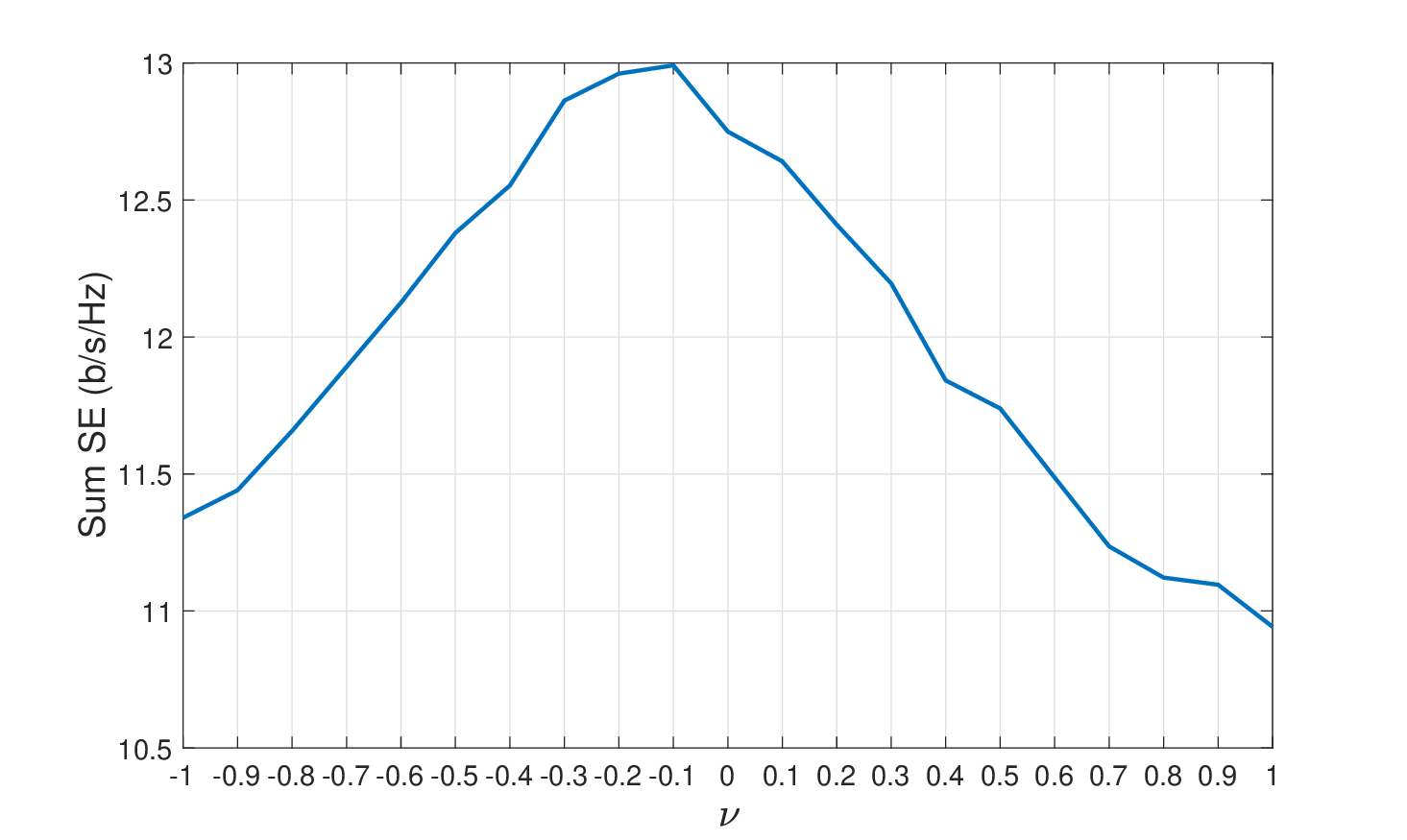}
\caption{Average sum spectral efficiency varying $\nu$. $M=64$, $P_{\mathrm{dl}}=33$ dBm, $Q_{\mathrm{ul}}=20$ dBm, 5 clusters , 8 users per cluster.}
\label{fig:Nu}
\end{figure}

\begin{figure}[t]
\centering
\includegraphics[width=\columnwidth]{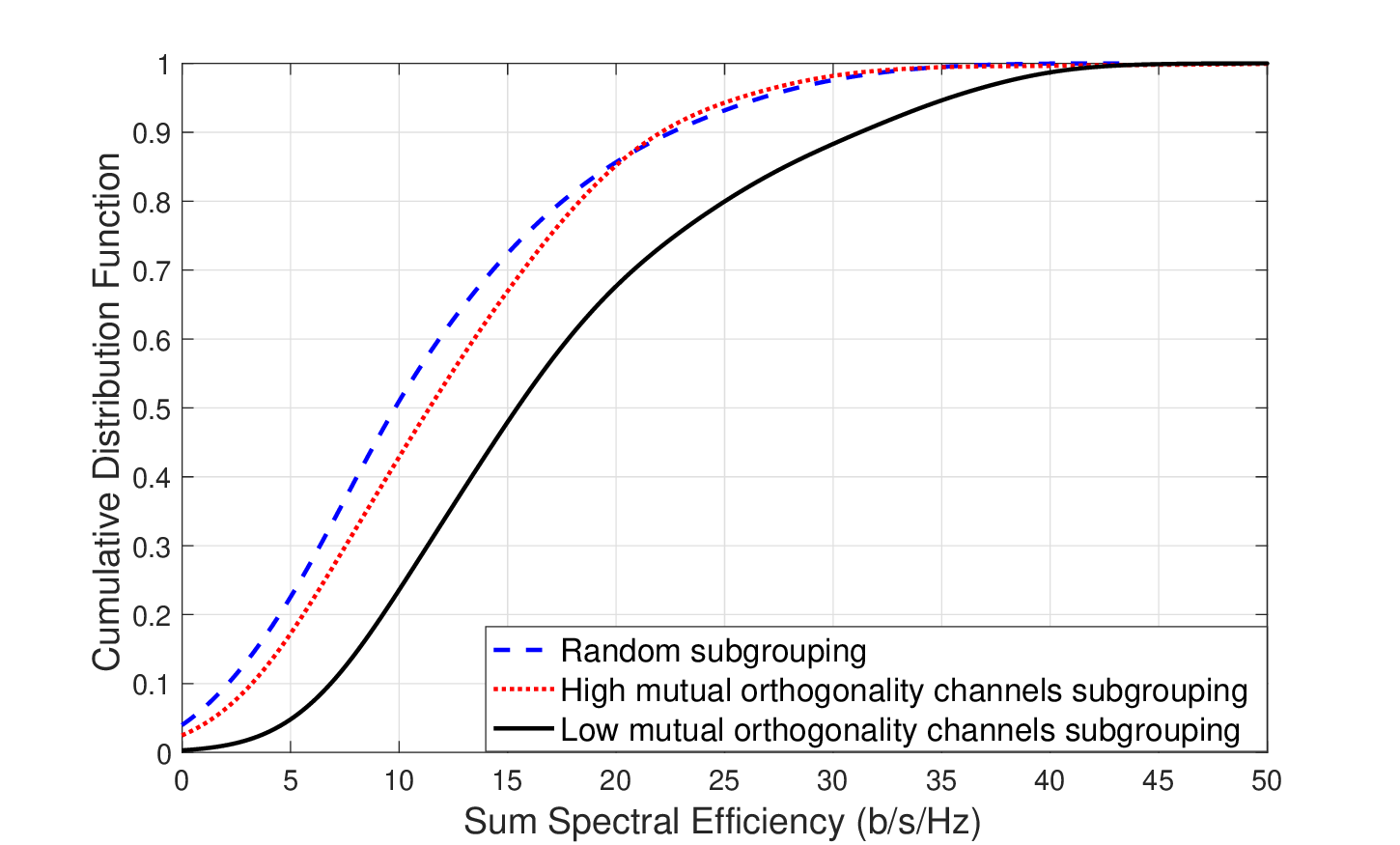}
\caption{CDF of the \gls{sse} using the MMF UL pilot and DL power control strategies, and different subgrouping criteria. $M=64$, $P_{\mathrm{dl}}=33$ dBm, $Q_{\mathrm{ul}}=20$ dBm, 7 clusters, 7 users per cluster, 7 subgroups.}
\label{fig:subgroup_criteria}
\end{figure}

In this subsection, we present a benchmark for the proposed subgrouping scheme, and assess the pre-determined DL power distribution given by Eq. (\ref{eq:p_g}) and employed in the UL pilot power control strategy. Figure \ref{fig:Nu} shows the average \gls{sse} achieved using our proposed UL pilot and DL power control strategies varying the parameter $\nu \in [-1,1]$. We observe that values of $\nu$ close to zero result in higher \gls{sse}, and the best performance are attained when employing $\nu=-0.1$. Therefore, this value is used in the simulations hereafter.
In Figure \ref{fig:subgroup_criteria}, we evaluate the performance of the proposed user subgrouping strategy. In these simulations, we assume $7$ clusters, each one including $7$ users, and consider three different subgrouping criteria. First, we consider $7$ subgroups in which users with low degree of mutual-orthogonality are grouped together (proposed scheme). Second, we consider $7$ subgroups in which users with high degree of mutual-orthogonality are grouped together. Third, we consider the case that the users are grouped randomly in $7$ subgroups. As the simulation results clearly show, the \gls{sse} achieved by the proposed subgrouping strategy is significantly larger than the other strategies. Specifically, subgrouping the users by their low mutual orthogonality results in a \gls{sse} higher than $6.85$ b/s/Hz with probability $90$\% (90\%-likely \gls{sse}), against $2.12$ and $3.22$ b/s/Hz achieved by the random subgrouping strategy, and the strategy that groups the users by their high mutual orthogonality, respectively. Hence, despite grouping together users with low mutual orthogonality increases the intra-subgroup interference and pilot contamination (recall that users of the same subgroup use the same pilot sequence), it is more convenient, in terms of \gls{sse}, to induce a higher spatial diversity between subgroups rather than between users of the same subgroup. Indeed, as the multicast transmission takes place on subgroup basis rather than on user basis, such a choice facilitates an effective spatial multiplexing of the subgroups.

\begin{figure}[t]
\centering
\includegraphics[width=\columnwidth]{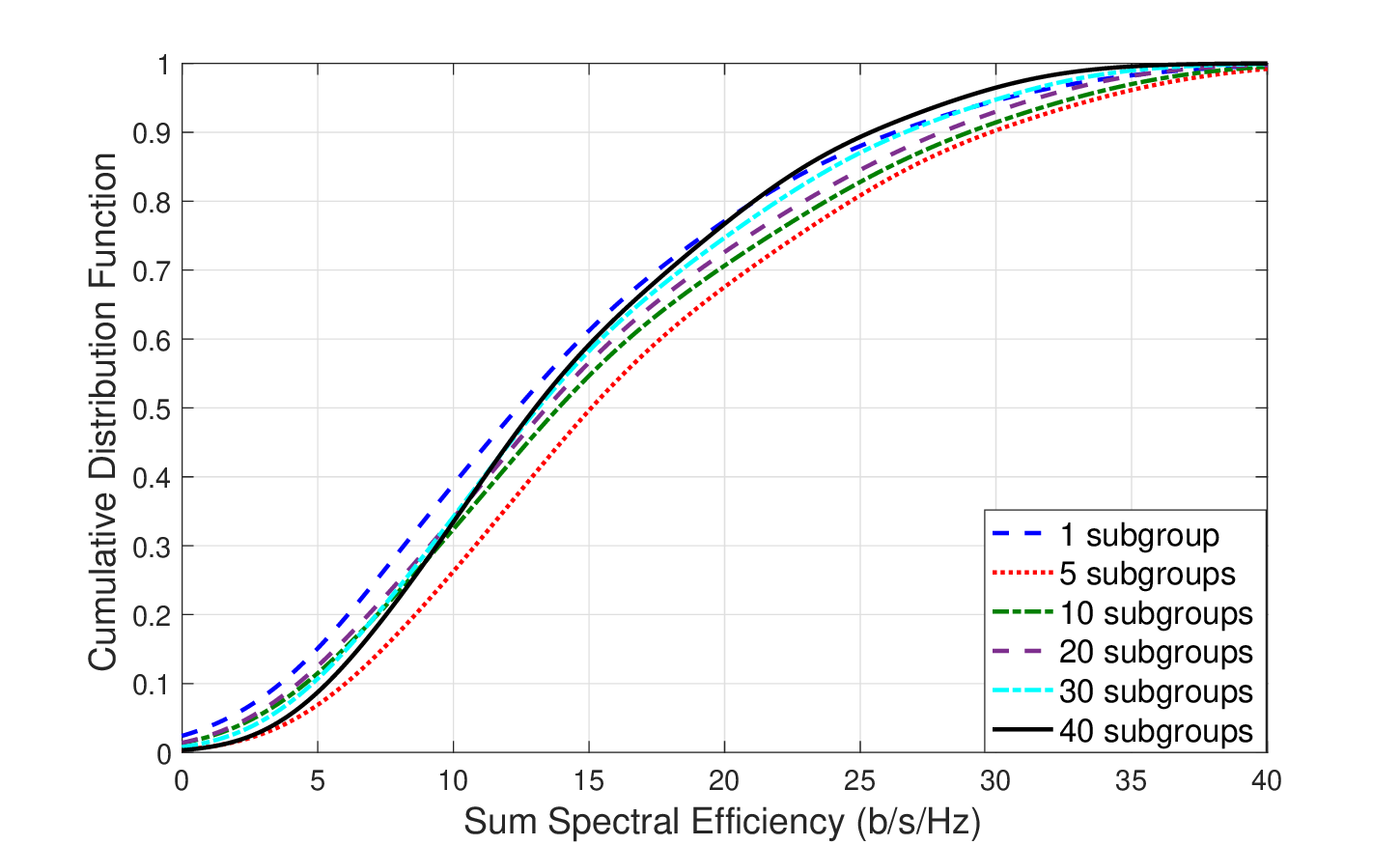}
\caption{CDF of the \gls{sse} using the MMF UL pilot and DL power control strategies and the proposed subgrouping criterion for different number of subgroups. $M=64$, $P_{\mathrm{dl}}=33$ dBm, $Q_{\mathrm{ul}}=20$ dBm, 5 clusters, 8 users per cluster.}
\label{fig:n_subgroups}
\end{figure}

Figure \ref{fig:n_subgroups} shows the cumulative distribution function (CDF) of the \gls{sse}, assuming the proposed subgrouping strategy and varying the number of subgroups in the multicast group. In these simulations, we consider $40$ users, geographically deployed in $5$ clusters (of $8$ users each) within the coverage area. Then, we evaluate the performance of the conventional multicast transmission, namely only one transmission to all the users, which form one multicast group. In addition, we consider the conventional massive MIMO operation with no multicasting, that is consisting in $40$ unicast transmissions. Finally, we consider our subroup multicast transmission strategy with different numbers of subgroups: $5$, $10$, $20$, and $30$. Simulation results reveal that the proposed strategy with $5$ multicast subgroups uniformly yields the highest \gls{sse}. Notice that, the optimal number of multicast subgroups is likely to reflect the number of deployed user clusters as our scheme captures the spatial similarities of users' channels\footnote{With the term \textit{cluster}, throughout all this paper, we indicate the deployment of a set of users with similar spatial characteristics, and the number of clusters is an input simulation parameter}.
For instance, we may observe that a 90\%-likely \gls{sse} of $6.02$ b/s/Hz is achieved by subgrouping the users in $5$ multicast subgroups, $5.33$ b/s/Hz is achieved by employing $40$ unicast transmissions, and $3.62$ b/s/Hz is attained by setting a conventional multicast transmission. Using $10$, $20$, and $30$ multicast subgroups results in a 90\%-likely \gls{sse} of $4.54$, $4.27$, and $4.81$ b/s/Hz, respectively. The performance improvement of the proposed scheme over any considered alternative is clearer in terms of median \gls{sse}. 

\begin{figure}[t]
\centering
\includegraphics[width=\columnwidth]{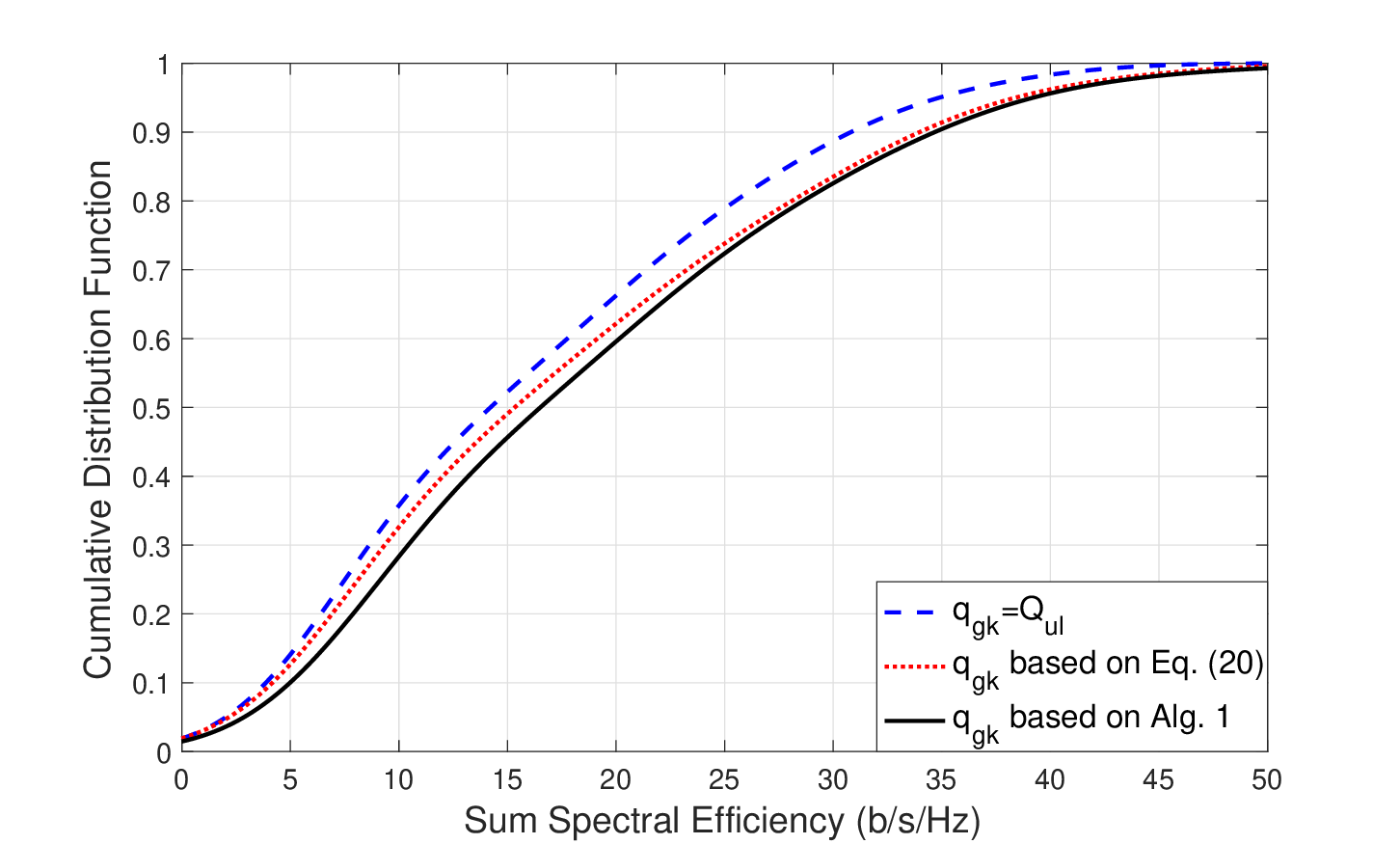}
\caption{CDF of the \gls{sse} using the proposed subgrouping criterion with MMF DL power control, and different UL pilot power allocations. $M=64$, $P_{\mathrm{dl}}=33$ dBm, $Q_{\mathrm{ul}}=20$ dBm, 3 clusters, 40 users per cluster, 3 subgroups.}
\label{fig:UL_strategy}
\end{figure}

In Figure \ref{fig:UL_strategy}, we present a performance comparison between our proposed UL pilot power control scheme and the closed-form power control solution given in Eq.~\eqref{eq:q_uncorr} by ~\cite{2018SadeghiTWC1}, which is optimal for uncorrelated Rayleigh fading channels. In this comparison, we also include a third option consisting of all the users transmitting their UL pilots with the available $Q_{ul}$ power transmission. In these simulations, we deploy $3$ clusters with $40$ users each and consider our subgroup multicast transmission operation with $3$ subgroups. Simulations results reveal a slight \gls{sse} improvement when using the proposed UL pilot MMF power control strategy. However, this should not be seen as a disappointing result, but rather a point in favor: our user subgrouping strategy and MMF DL power allocation counteract (or capitalize on) the spatial correlations among the users' channels letting the uplink pilot power allocation less relevant in maximizing the multicast \gls{sse}. Hence, either using the power allocation in Eq.~\eqref{eq:q_uncorr} or simply full pilot power transmission is equivalently effective with our subgroup multicast transmission strategy along with the proposed MMF DL power allocation. As a result, a low-complexity UL pilot power allocation can be conveniently performed.
Specifically, we observe that the proposed UL pilot power allocations achieve a 90\%-likely \gls{sse} of $4.98$ b/s/Hz, while $4.17$ and $3.87$ b/s/Hz are the 90\%-likely \gls{sse} achieved by for the UL pilot power control strategy in~\cite{2018SadeghiTWC1} and using $Q_{ul}$ for all the users, respectively. This interesting result is even more relevant when the number of multicast users per deployed cluster decreases, as it will be shown in the following subsections. Only in scenarios with a very large number of multicast users per cluster (i.e., more than $100$ multicast users in a deployed cluster), the development of sophisticated UL pilot power control schemes can result in relevant \gls{sse} benefits.
\begin{figure}[t]
\centering
\includegraphics[width=\columnwidth]{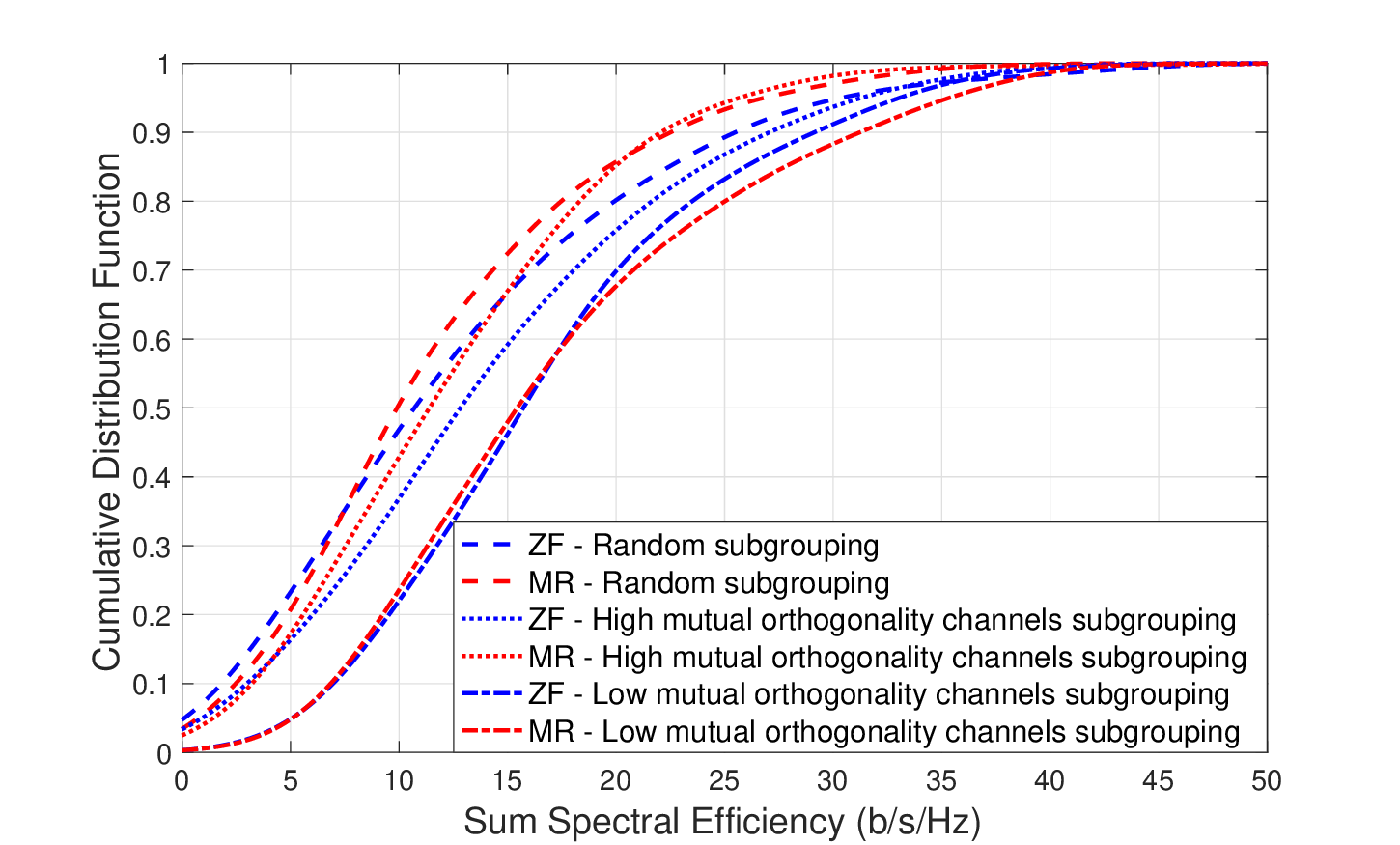} %
\caption{CDF of the \gls{sse} using the proposed subgrouping criterion with MMF UL/DL power control, and different precoding techniques and number of users per cluster. $M=64$, $P_{\mathrm{dl}}=33$ dBm, $Q_{\mathrm{ul}}=20$ dBm, 7 clusters, 7 users per cluster, 7 subgroups, 7 users per subgroup.}
\label{fig:MRvsZF}
\end{figure}
\begin{figure*}[!t]
\subfloat[$1$ cluster of $40$ users per cluster]
{\includegraphics[width=0.33\textwidth]
{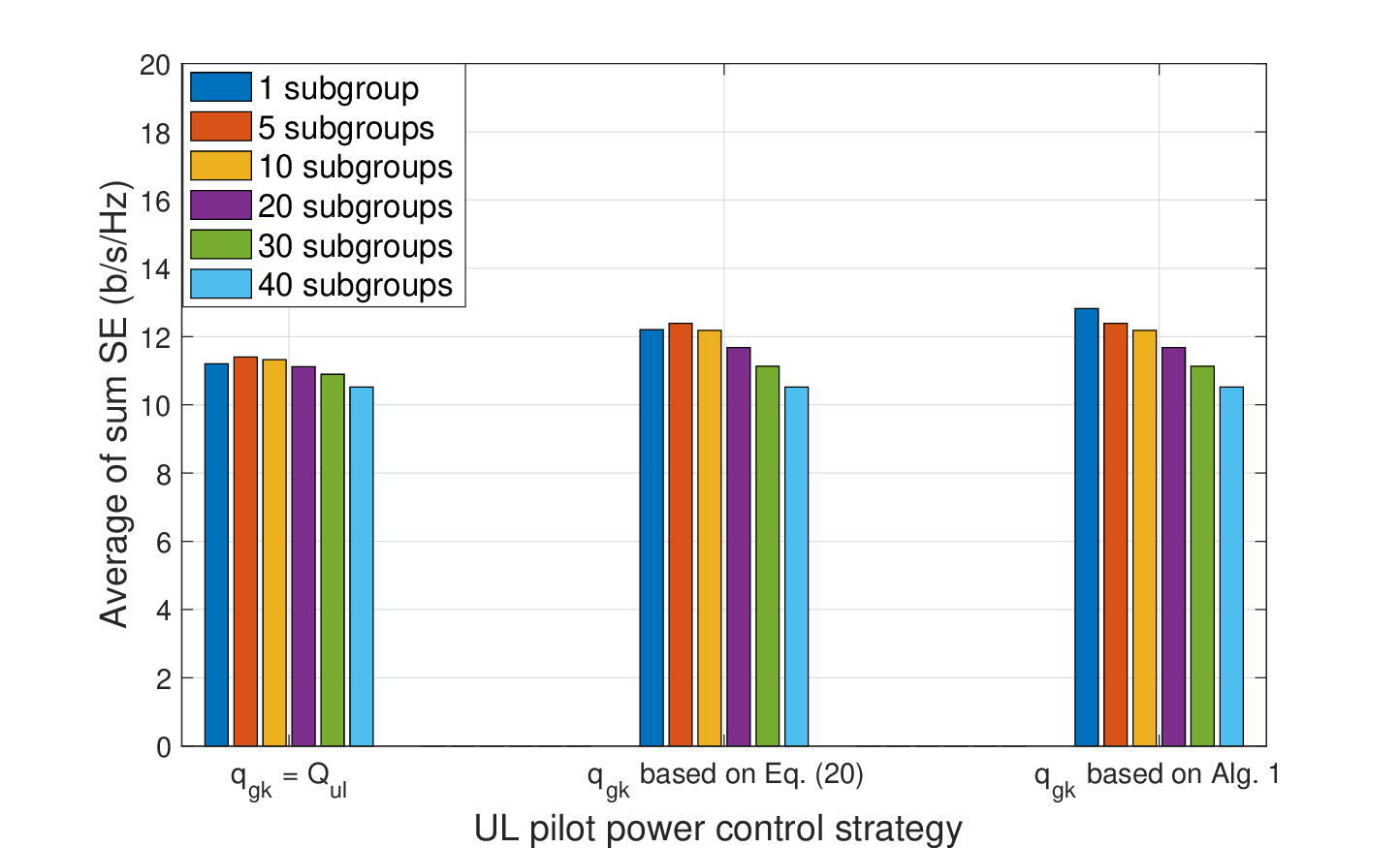}\label{fig:SumSE1x40}}
\subfloat[$2$ clusters of $20$ users per cluster]
{\includegraphics[width=0.33\textwidth]
{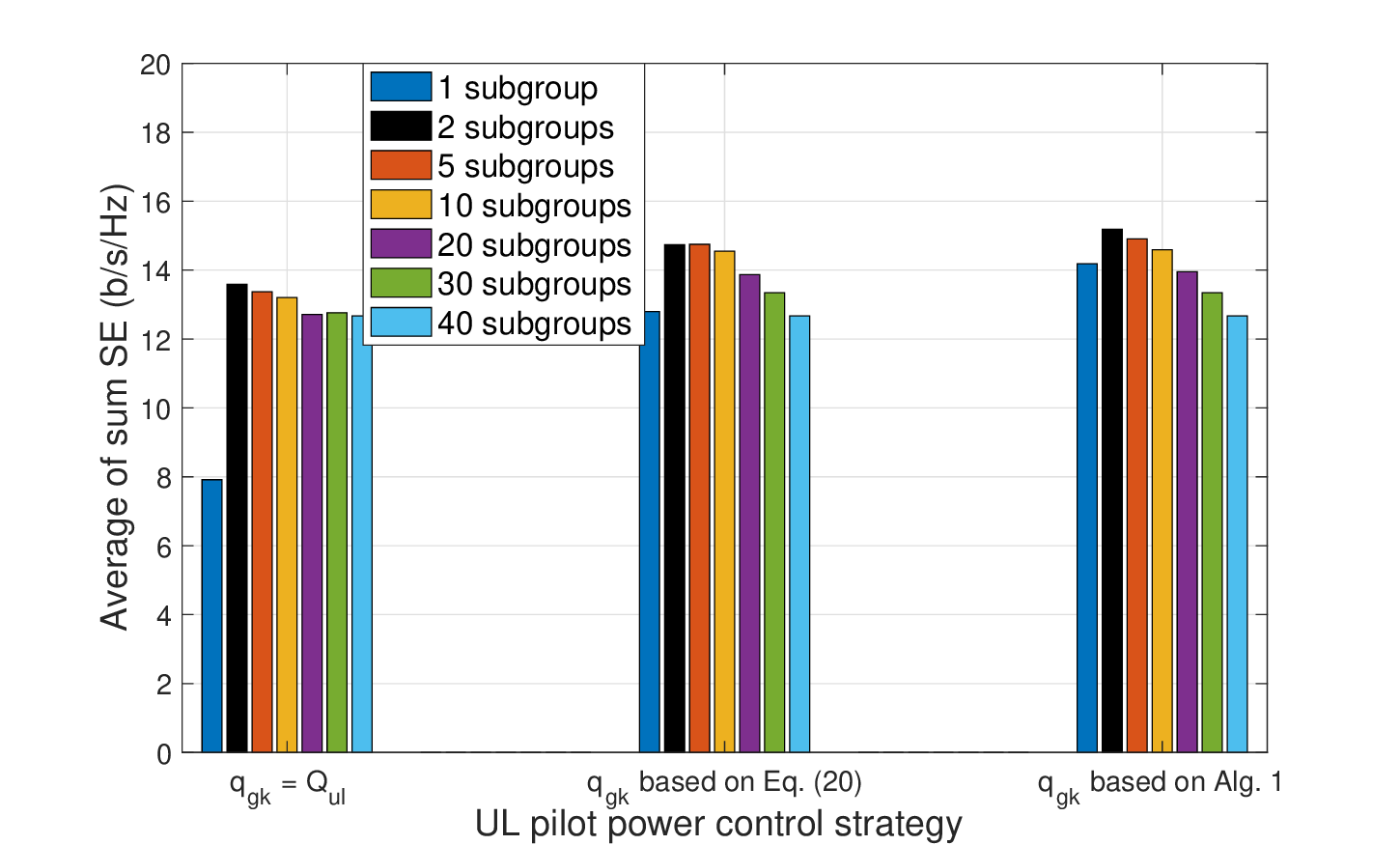}\label{fig:SumSE2x20}}
\subfloat[$5$ clusters of $8$ users per cluster]
{\includegraphics[width=0.33\textwidth]
{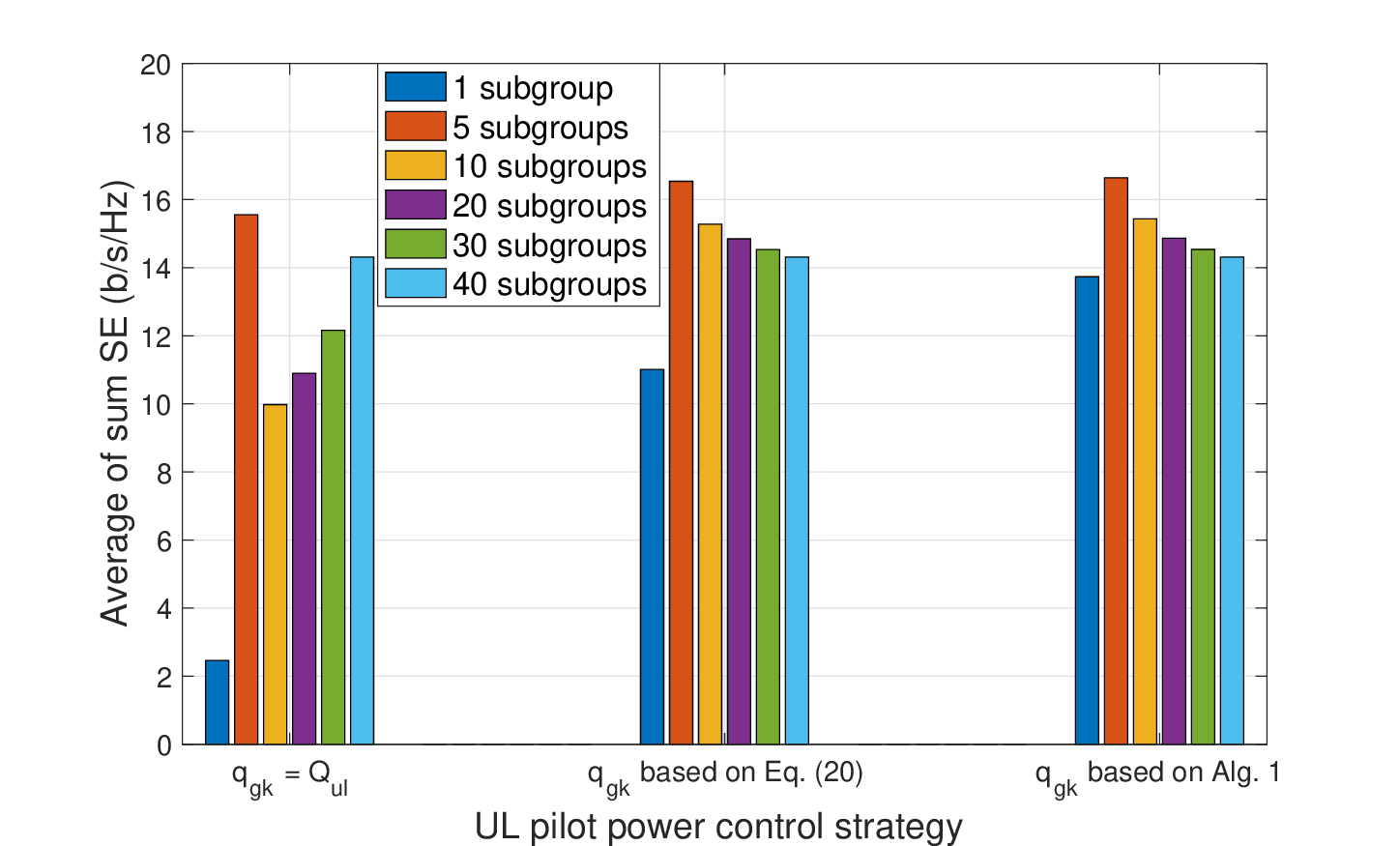}\label{fig:SumSE5x8a}}\\
\subfloat[$8$ clusters of $5$ users per cluster]
{\includegraphics[width=0.33\textwidth]
{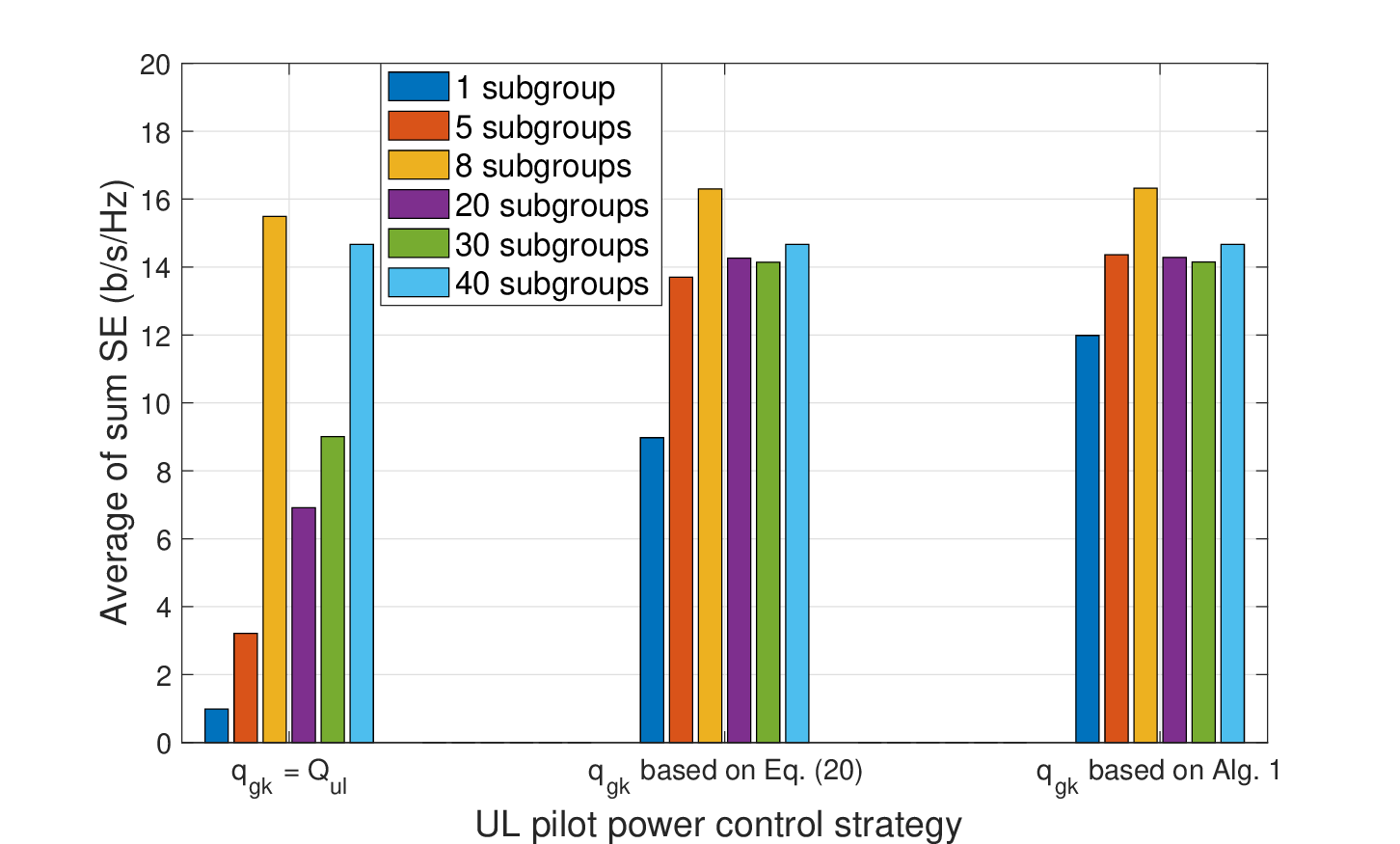}\label{fig:SumSE8x5}}
\subfloat[$20$ clusters of $2$ users per cluster]
{\includegraphics[width=0.33\textwidth]
{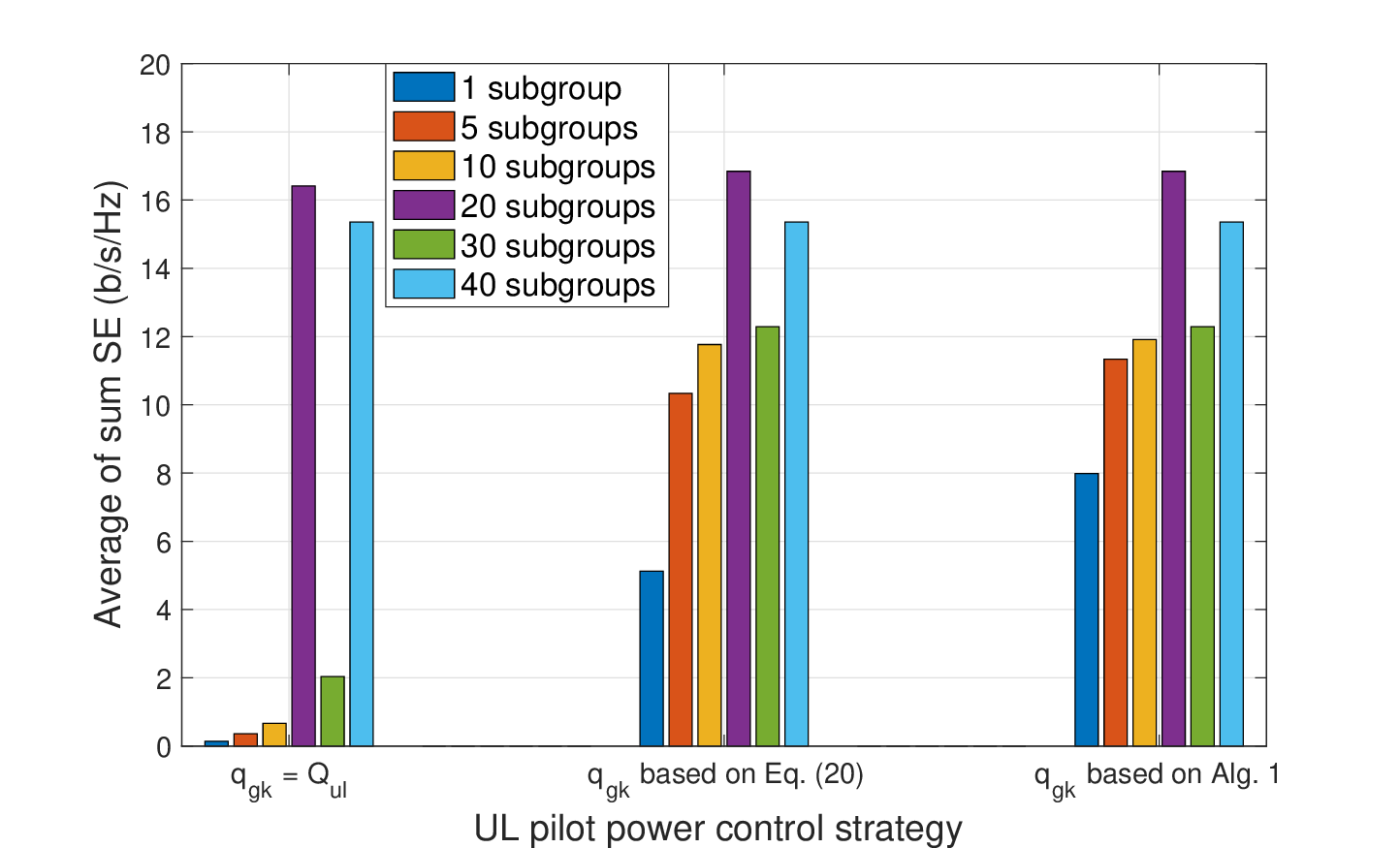}\label{fig:SumSE20x2}}
\subfloat[$40$ clusters of $1$ user per cluster]
{\includegraphics[width=0.33\textwidth]
{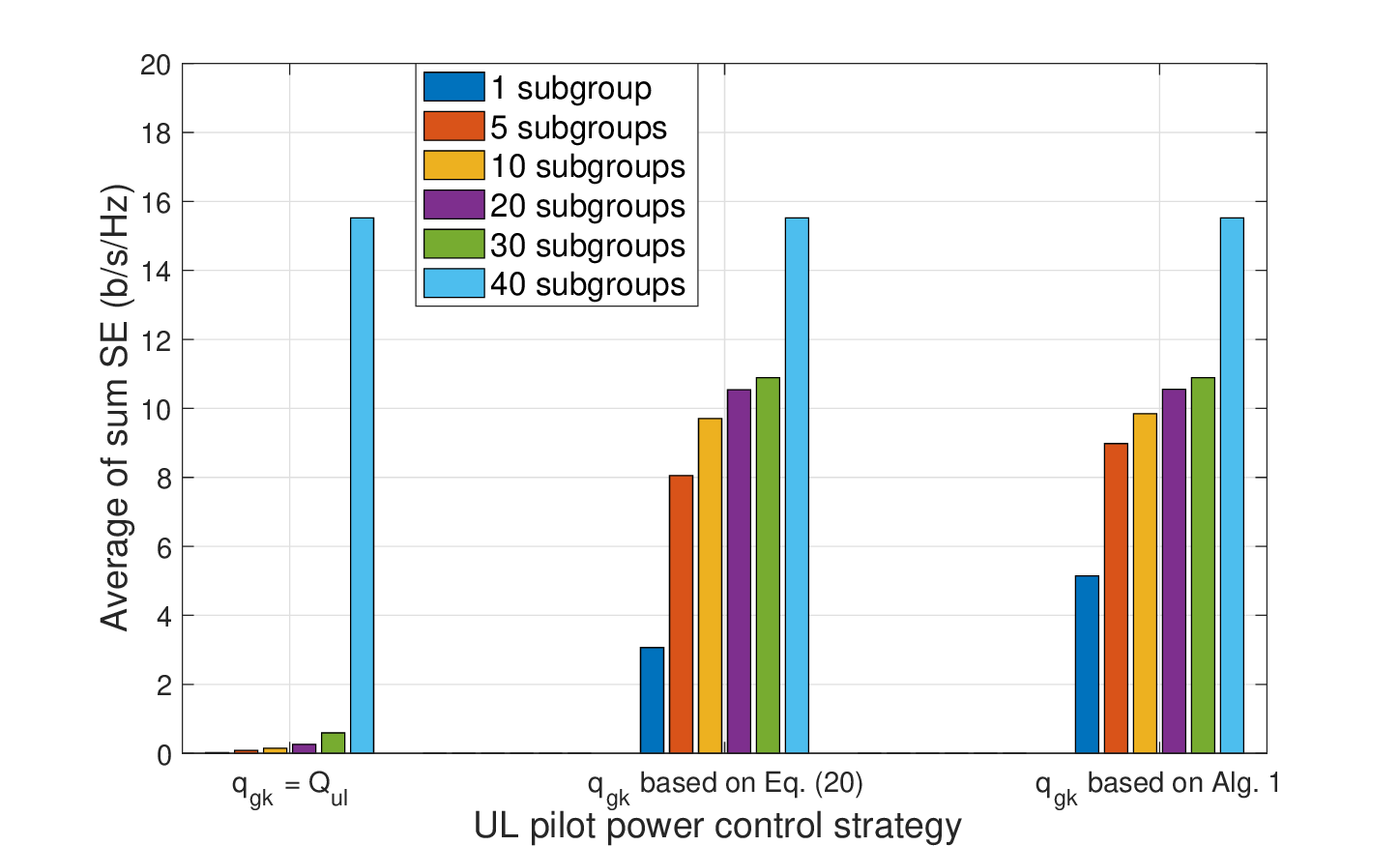}\label{fig:SumSE40x1}}
\caption{{Average \gls{sse} using the MMF DL power control, different UL pilot power control strategies, and different number of multicast subgroups. $M=64$, $P_{\mathrm{dl}}=33$ dBm, $Q_{\mathrm{ul}}=20$ dBm, $K=40$ with different number of clusters and users per cluster.}}
\label{fig:number_clusters}
\end{figure*}

Figure \ref{fig:MRvsZF} compares the \gls{sse} achieved by using either MR transmission or ZF precoding towards each multicast subgroup. In these simulations, we consider $7$ clusters with $7$ users each, and apply the three user subgrouping strategies as in the simulations of Figure~\ref{fig:subgroup_criteria}.
Again, we may observe a relevant result: our grouping strategy consisting in grouping together the users with similar spatial characteristics (i.e., channels with low mutual orthogonality) is able to fill the gap between MR and ZF essentially enabling multicast transmissions free of interference. MR even outperforms ZF at higher percentiles as ZF uses some degrees of freedom to suppress the interference in some orthogonal spatial directions, thereby reducing the effective array gain. Our subgrouping strategy has therefore a great ability to make the inter-subgroup interference negligible, unlike other subgrouping strategy. Importantly, this allows an efficient use of low-complexity precoding schemes such us MR transmission.   
Conversely, ZF precoding with either random or high degree of mutual-orthogonality subgrouping provides higher \gls{sse} than its MR counterpart, especially at high percentiles, but it is still far from performing as well as our proposed scheme. This further confirms that other user subgrouping strategies would severely suffer of both inter-subgroup interference and pilot contamination, which cannot be entirely suppressed even by using ZF precoding.

\subsection{Effects of the number of users per cluster}

In this subsection, we assess the impact of the number of users per cluster on the performance of the proposed subgrouping strategy.

Figure \ref{fig:SumSE1x40} shows the results when all the users are deployed in only one cluster. The average \gls{sse}, $12.82$ b/s/Hz, is achieved using only one subgroup with the UL pilot power control strategy proposed in Algorithm 1. Creating $5$ subgroups with either the UL pilot power control in Algorithm 1 or in~\cite{2018SadeghiTWC1} provides an average \gls{sse} of $12.39$ b/s/Hz. Figure \ref{fig:SumSE2x20} illustrates the case of deploying the users in $2$ location clusters. We see that creating $2$ subgroups with our proposed UL pilot power control strategy provides the highest \gls{sse}, $15.18$ b/s/Hz. In Figure \ref{fig:SumSE5x8a}, our subgrouping and UL pilot power control strategy achieves the highest average sum SE with $16.64$ b/s/Hz. While, Figures \labelcref{fig:SumSE8x5,fig:SumSE20x2,fig:SumSE40x1} illustrate the results using $8$, $20$, and $40$ clusters of users, respectively. We notice that using $8$, $20$, and $40$ subgroups with the Algorithm 1 is always the best solution providing \gls{sse} of $16.32$, $16.84$, and $15.52$ b/s/Hz, respectively. 

These results confirm, as mentioned earlier, that the optimal number of subgroups reflects the number of location clusters deployed. Hence, once the user clusters are identified upon the knowledge of the large-scale fading quantities, determining the pilots' assignment and the number of multicast transmissions (i.e., determining the multicast subgroups) is immediate. Furthermore, once the optimal number of subgroups is established, any UL pilot power control strategy is effective. Nevertheless, when the number of users per subgroup is increased (i.e., $20$ or $40$ users per subgroup in Figs 9b and 9a, respectively), the benefits of employing the proposed UL pilot power control strategy can become relevant.
\begin{figure*}[!t]
\subfloat[$5$ clusters of $4$ users per cluster]
{\includegraphics[width=0.33\textwidth]
{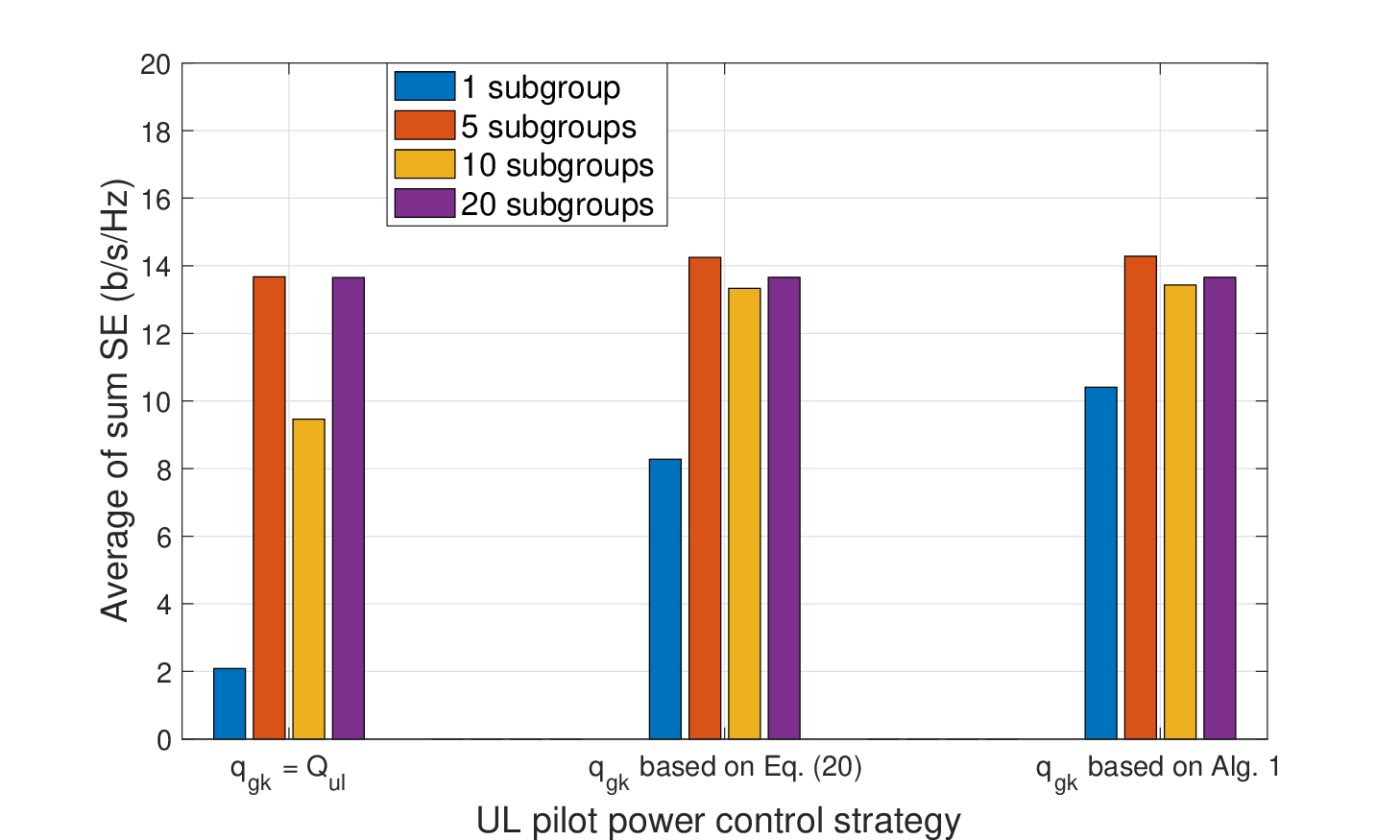}\label{fig:SumSE5x4}}
\subfloat[$5$ clusters of $8$ users per cluster]
{\includegraphics[width=0.33\textwidth]
{SumSE5x8usersMMF_MR500.eps}\label{fig:SumSE5x8_b}}
\subfloat[$5$ clusters of $20$ users per cluster]
{\includegraphics[width=0.33\textwidth]
{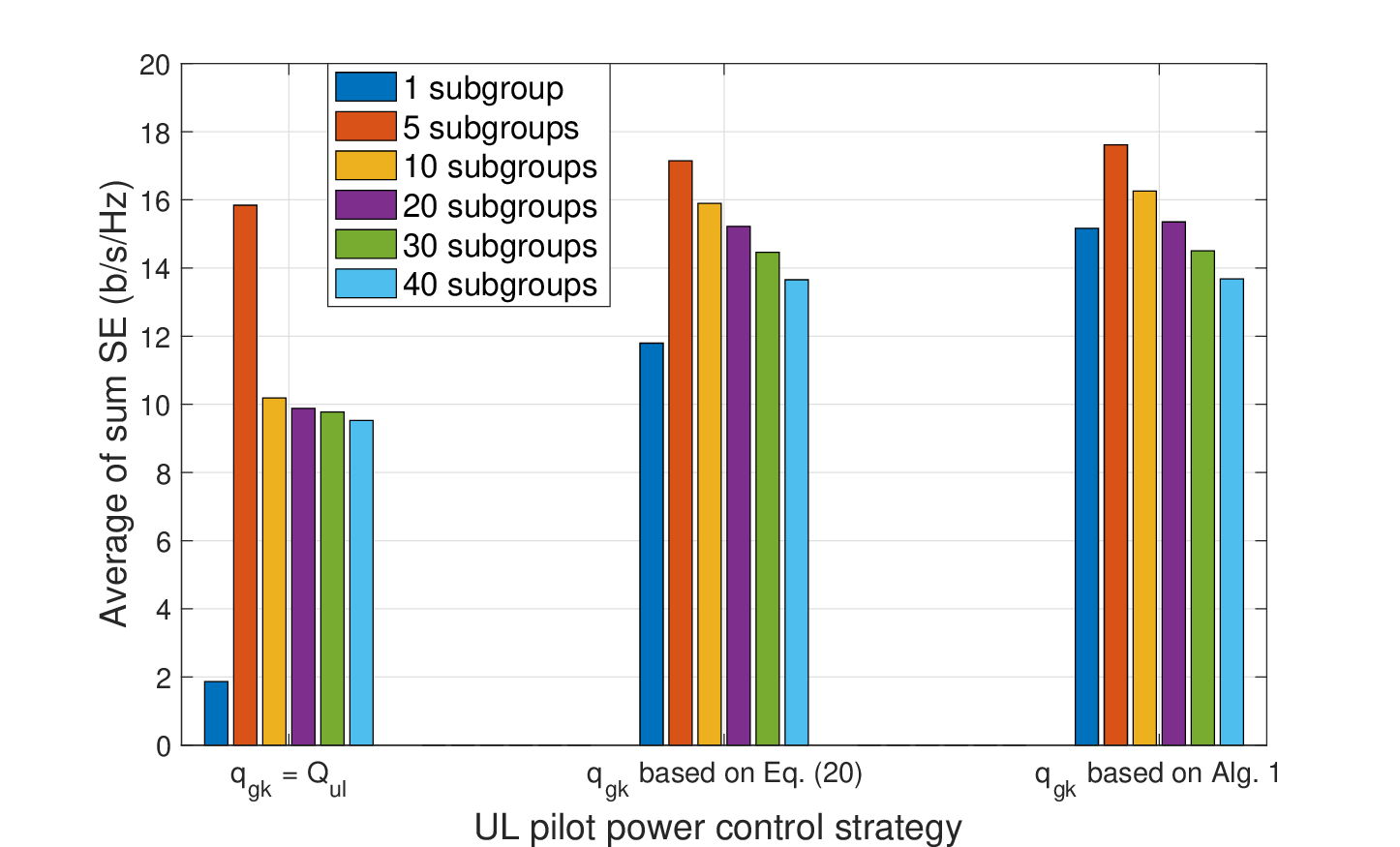}\label{fig:SumSE5x20}}\\
\caption{{Average sum spectral efficiency using the MMF DL power control, different UL pilot power control strategies, and different number of multicast subgroups. $M=64$, $P_{\mathrm{dl}}=33$ dBm, $Q_{\mathrm{ul}}=20$ dBm, 5 clusters with different number of users per cluster. }}
\label{fig:number_users}
\end{figure*}

\subsection{Effects of the number of multicast users}

In this subsection, we evaluate the impact of the total number of multicast users on the performance of the proposed subgrouping and UL pilot power control strategies.

In Figure \ref{fig:number_users}, we show the average \gls{sse} for an increasing number of multicast users, while the number of location clusters is fixed to $5$ clusters. Specifically, we consider the configurations with $4$, $8$, and $20$ users per cluster. 
Figures \ref{fig:SumSE5x4}--\ref{fig:SumSE5x20} point out that the highest average  \gls{sse} is obtained whenever the number of multicast transmissions is equal to the number of clusters regardless of the UL pilot power control schemes. In any other case, our proposed MMF UL pilot power allocation described in Algorithm 1 performs the best by far, especially with the traditional multicast transmission (only $1$ subgroup) by ignoring the spatial correlation. This trend does not vary as the number of users per subgroup increases, even though the use of our proposed  MMF UL pilot power control with the subgrouping strategy results in more relevant benefits with a larger number of users per subgroup. However, the main take-home message of these results is that even increasing the number of users per subgroup, hence increasing the level of intra-subgroup interference and pilot contamination, no sophisticated UL pilot power control strategies are needed as long as the users are grouped by a low degree of channel mutual orthogonality and the number of subgroups is properly identified.  Finally, despite the increase of the overall level of interference in the system as the number of users grows, we may still appreciate a slight \gls{sse} improvement whenever multicasting is adopted. Indeed, a performance degradation is observed only with conventional massive MIMO operation, namely only with unicast transmission, e.g., see the \gls{sse} degradation from the 40-subgroups setup in \ref{fig:SumSE5x8_b} to the 40-subgroups setup in \ref{fig:SumSE5x20}. This conclusion suggests that our proposed subgrouping multicasting scheme enables more efficient utilization of the radio resources. 

\subsection{Effects of the cluster radii}

\begin{figure*}[!t]
\subfloat[Cluster radius = 2.5 m]
{\includegraphics[width=0.33\textwidth]
{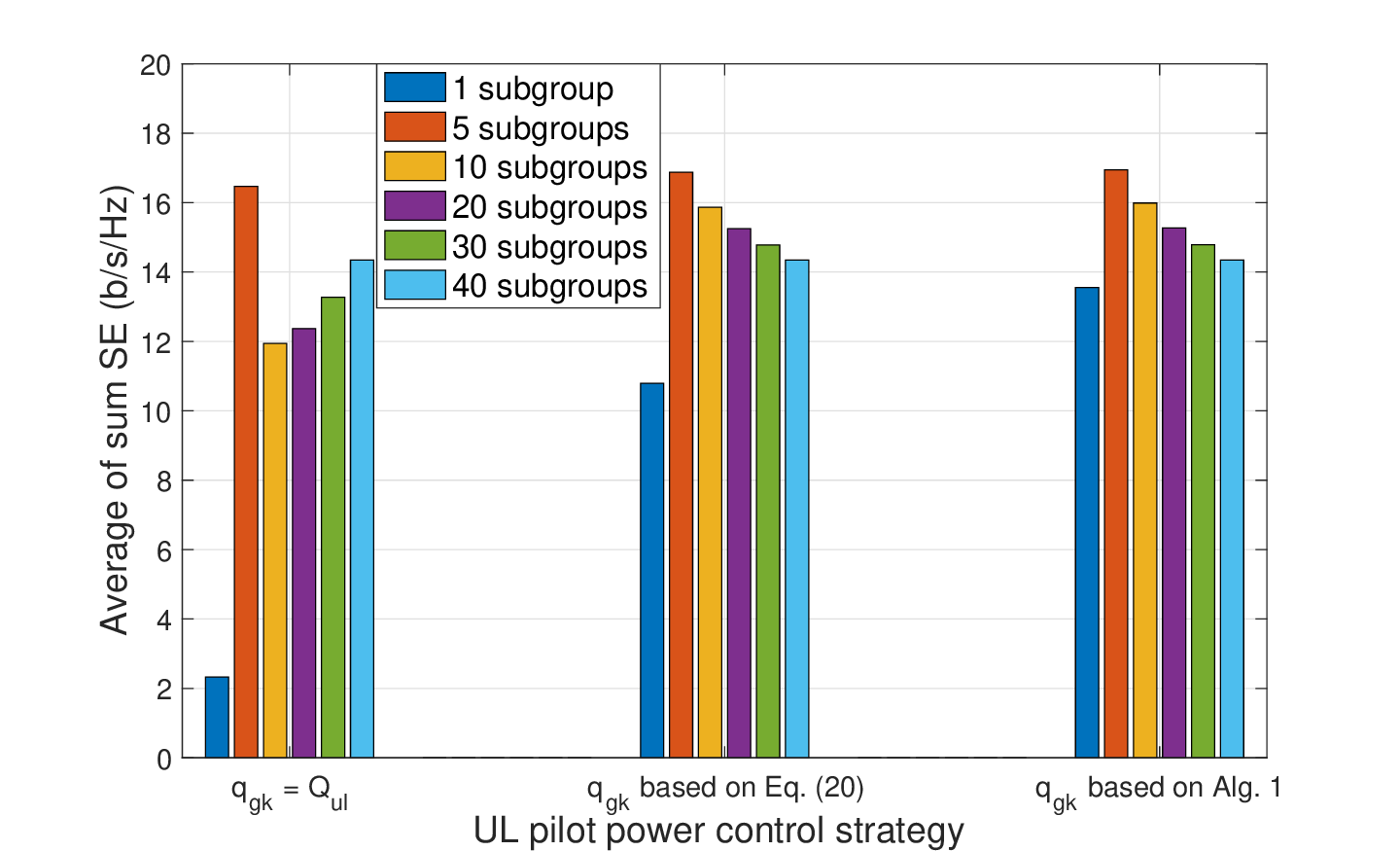}\label{fig:SumSE2.5m}}
\subfloat[Cluster radius = 5 m]
{\includegraphics[width=0.33\textwidth]
{SumSE5x8usersMMF_MR500.eps}\label{fig:SumSE5m}}
\subfloat[Cluster radius = 15 m]
{\includegraphics[width=0.33\textwidth]
{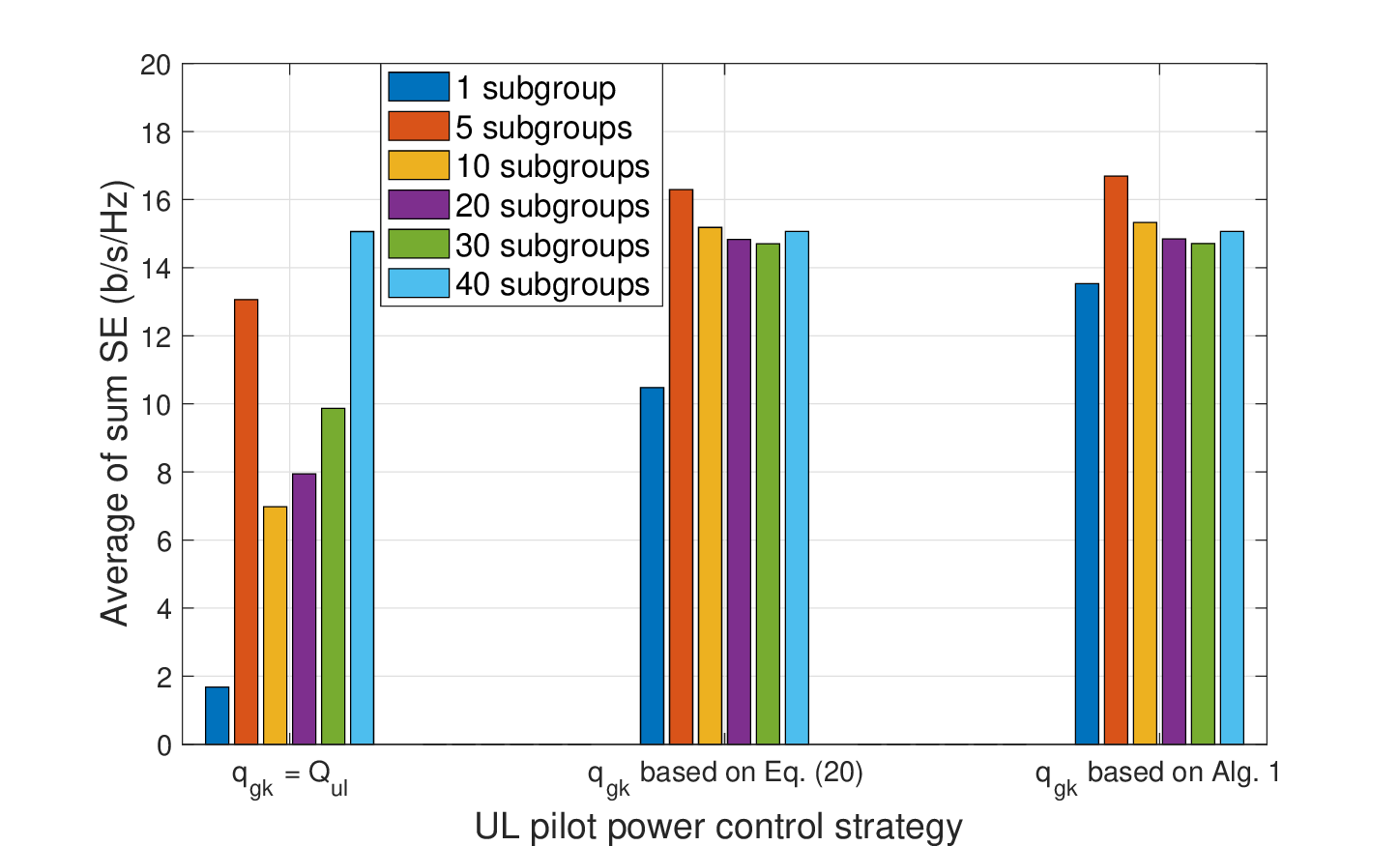}\label{fig:SumSE15m}}\\
\caption{{Average sum spectral efficiency using the MMF DL power control and different UL pilot power control strategies, number of multicast subgroups, and cluster radius. $M=64$, $P_{\mathrm{dl}}=33$ dBm, $Q_{\mathrm{ul}}=20$ dBm, 5 clusters, 8 users per cluster. }}
\label{fig:cluster_radius}
\end{figure*}

In this subsection, we consider the impact of the cluster radii on the performance of the proposed subgrouping and UL pilot power control strategies. Notice that a smaller cluster radius leads to group users with very low channel mutual orthogonality. Importantly, in the case of narrow clusters, users with nearly-similar spatial characteristic may be grouped in different subgroups causing a non-negligible inter-subgroup interference. Conversely, when increasing the cluster radius, users with significant spatial dissimilarities may be grouped together and, as a result, preventing an efficient utilization of the radio resources. This insight is however confirmed, in Figure \ref{fig:cluster_radius}, only when using UL pilot full power transmission, while the MMF UL pilot power allocations defined in~\eqref{eq:q_uncorr} and Algorithm 1 are quite robust against an inaccurate choice of the cluster radius. Indeed, while it is obvious that, in the considered scenario, a cluster with radius 2.5 meters is proper for the UL pilot full power transmission strategy, the performance achieved by the MMF UL pilot power allocations keep being constant as the cluster radius varies, especially when the number of subgroups equals the number of clusters.
%

\begin{figure*}[!t]
\subfloat[$P_{dl} = 33$ dBm, $Q_{ul} = 20$ dBm]
{\includegraphics[width=0.33\textwidth]
{SumSE5x8usersMMF_MR500.eps}\label{fig:SumSE2W01W}}
\subfloat[$P_{dl} = 46$ dBm, $Q_{ul} = 20$ dBm]
{\includegraphics[width=0.33\textwidth]
{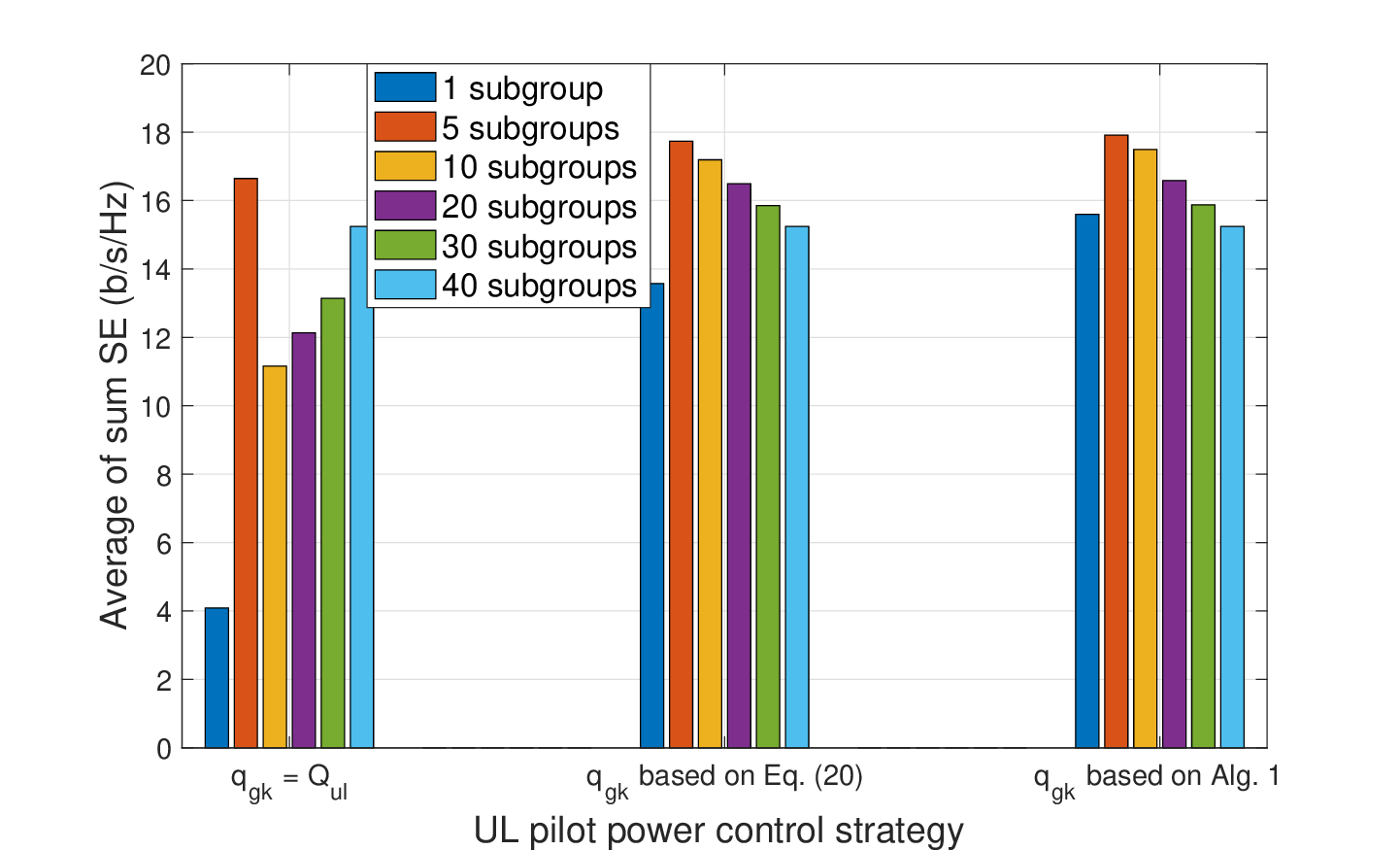}\label{fig:SumSE40W01W}}
\subfloat[$P_{dl} = 33$ dBm, $Q_{ul} = 30$ dBm]
{\includegraphics[width=0.33\textwidth]
{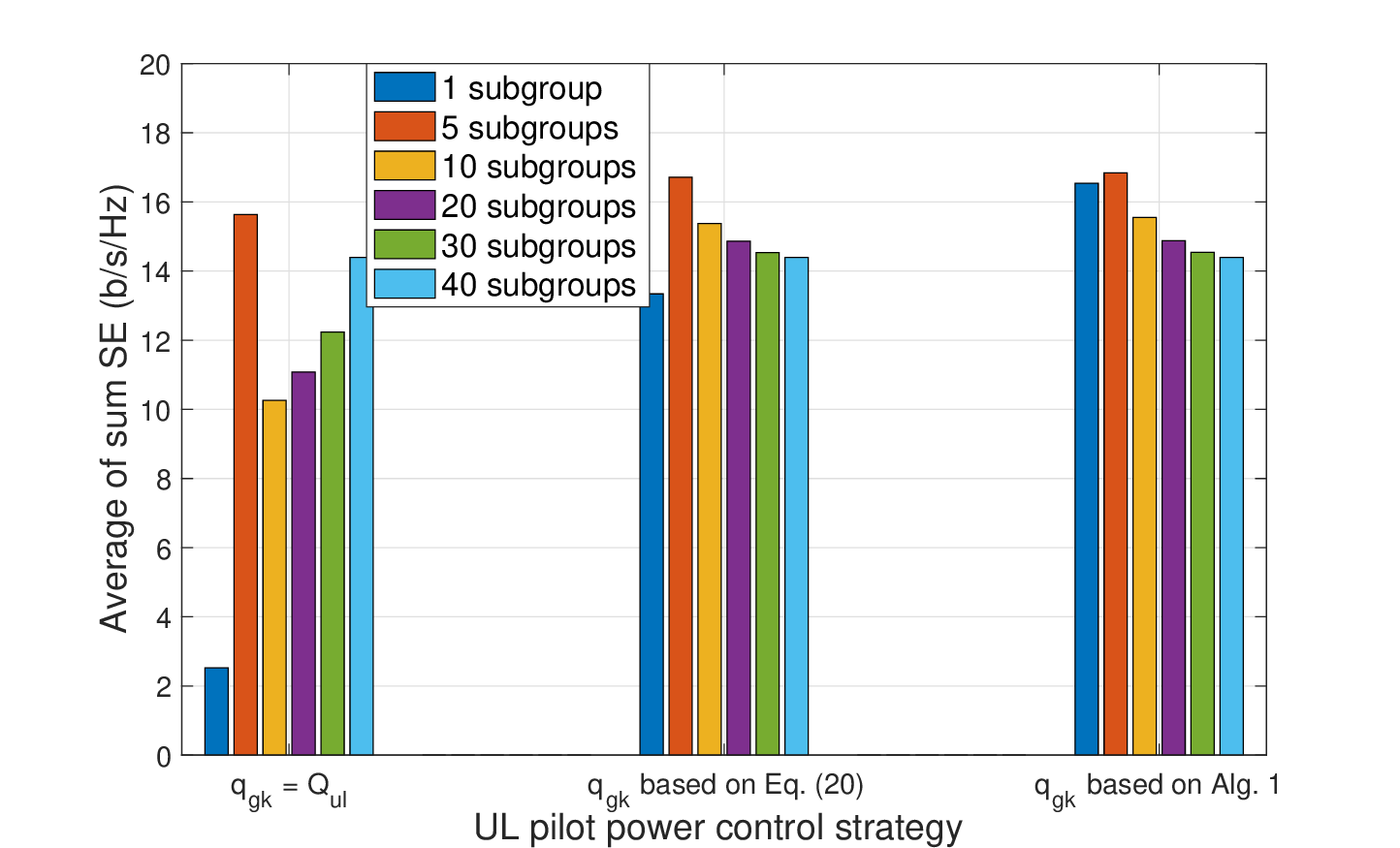}\label{fig:SumSE2W1W}}\\
\caption{{Average of sum spectral efficiency using the MMF DL power control and different UL pilot power control strategies, different number of multicast subgroups, and UL and DL power budgets. $M=64$, $P_{\mathrm{dl}}=33$ dBm, $Q_{\mathrm{ul}}=20$ dBm, 5 clusters, 8 users per cluster.}}
\label{fig:UL_DL_power}
\end{figure*}

\subsection{Effects of the UL and DL power budget}

In this subsection, we analyze the effect of the UL pilot and DL power budgets on the performance of the proposed subgrouping and UL pilot power control strategies. 

In Fig. \ref{fig:UL_DL_power}, we present the average \gls{sse} of $40$ users located in $5$ clusters when we increase the UL pilot or DL power budget. Examining \Cref{fig:SumSE2W01W,fig:SumSE40W01W}, we observe that the boost of the DL power budget from $33$ to $46$ dBm results in an increase of the \gls{sse} from $16.64$ to $17.91$ b/s/Hz. Hence, despite the increase of the overall level of interference when increasing the transmit powers and MR precoding is adopted, the combined use of MMF DL power control and the subgrouping multicast strategy still enables to increase the \gls{sse}.  
Observing \Cref{fig:SumSE2W01W,fig:SumSE2W1W}, we notice that increasing the UL pilot power budget from $20$ to $30$ dBm does not provide a significant change on the average \gls{sse} achieved by the optimal subgroup configuration, i.e., $5$ subgroups. As a remark, a noticeable \gls{sse} improvement is instead observed if adopting the conventional multicast transmission (i.e., bar plot corresponding to the 1 subgroup setup). Hence, even though increasing the UL pilot or DL power budget can improve the results of the non-optimal subgrouping configurations, the proposed subgrouping scheme is able to achieve excellent \gls{sse} with modest power budgets.

\subsection{Effects of the number of \gls{bs} antennas}

\begin{figure*}[!t]
\subfloat[BS transmit antennas = 64]
{\includegraphics[width=0.33\textwidth]
{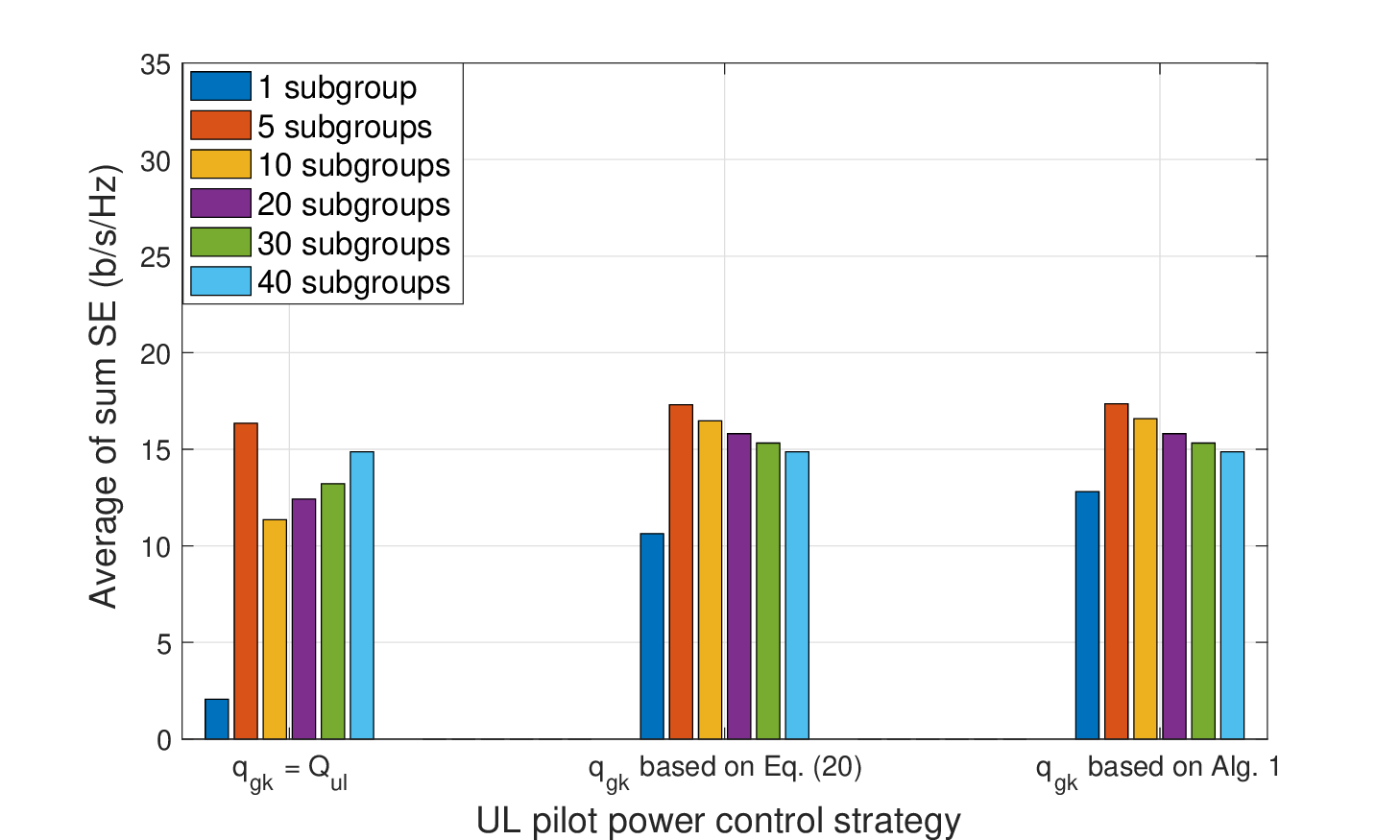}\label{fig:SumSE64ant}}
\subfloat[BS transmit antennas = 128]
{\includegraphics[width=0.33\textwidth]
{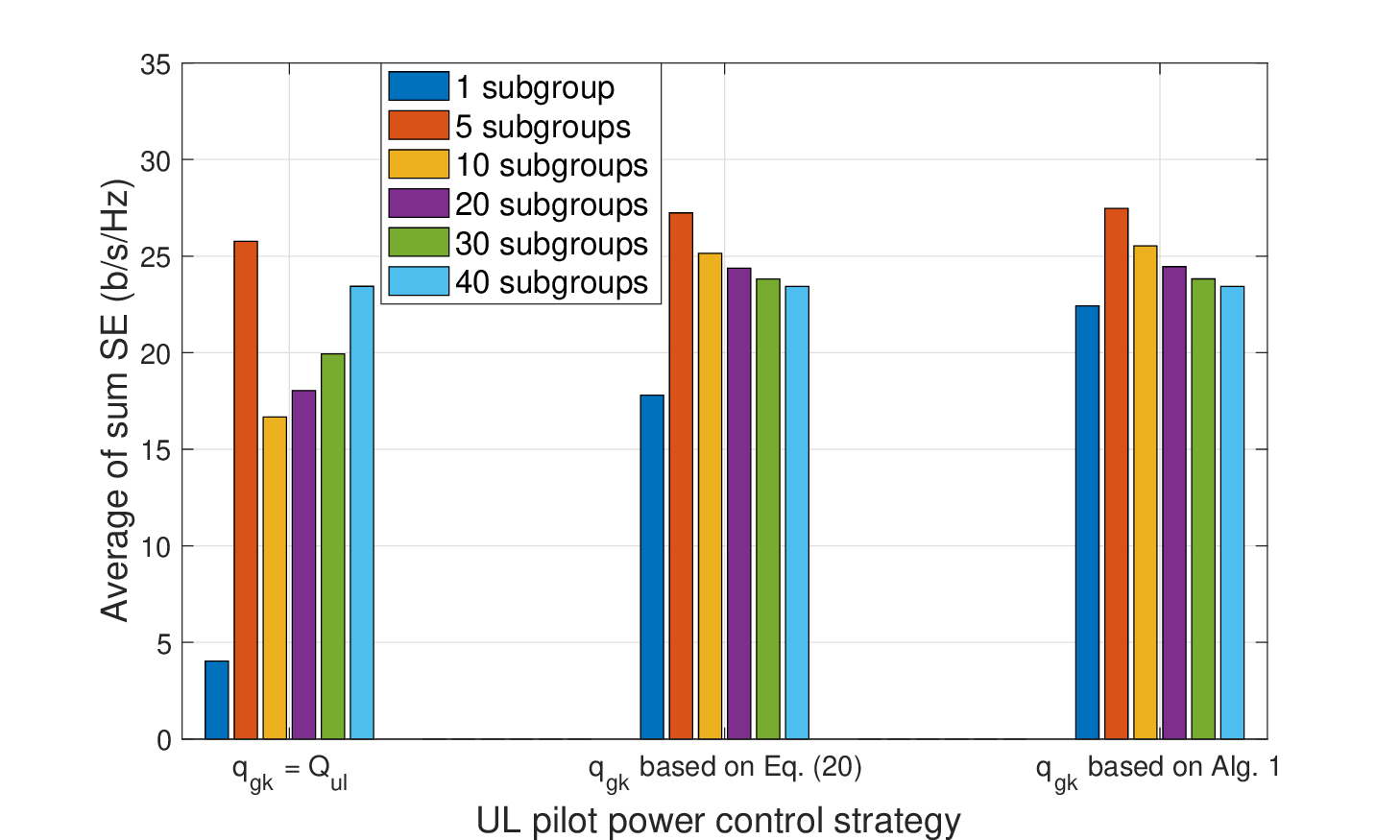}\label{fig:SumSE128ant}}
\subfloat[BS transmit antennas = 192]
{\includegraphics[width=0.33\textwidth]
{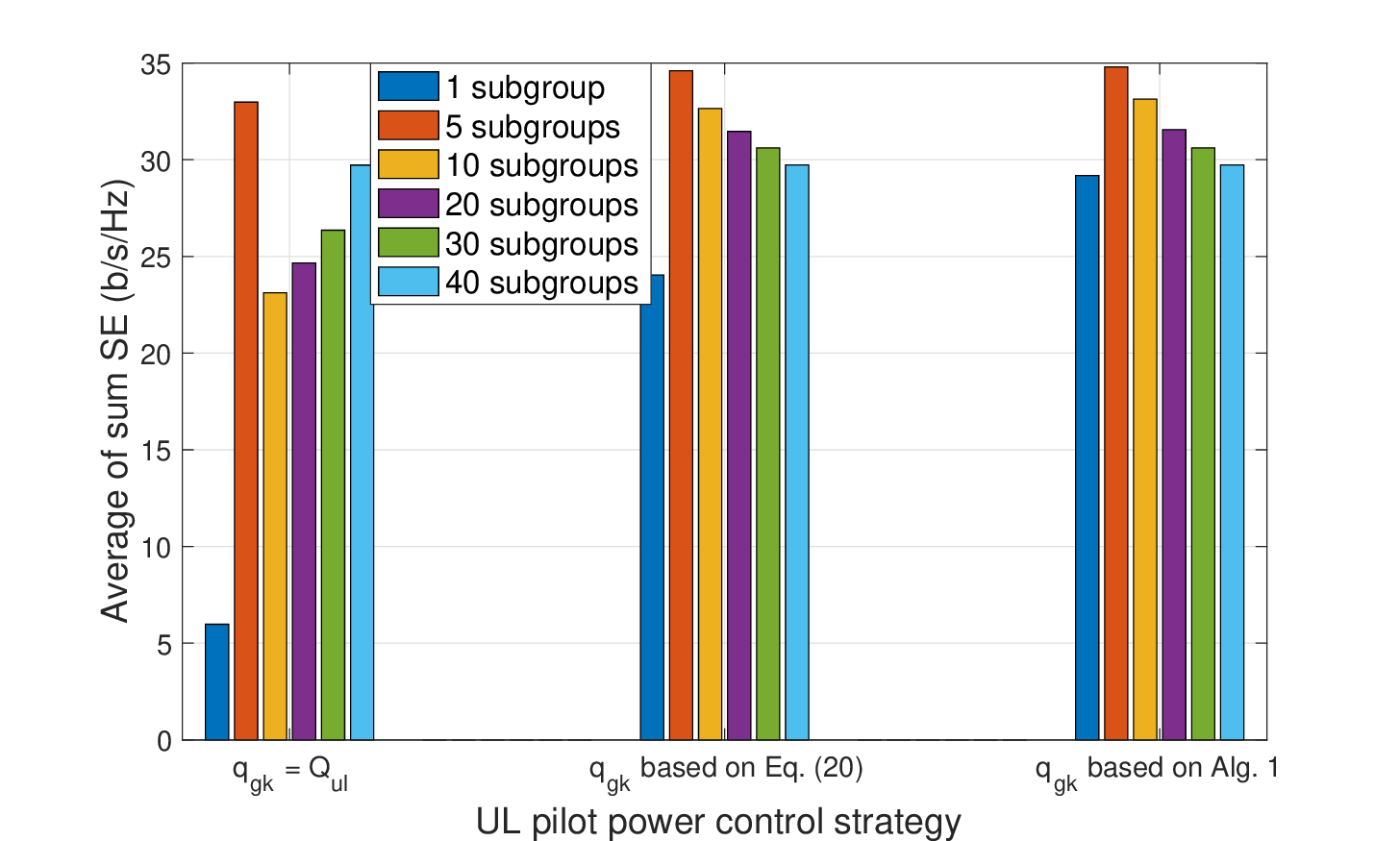}\label{fig:SumSE192ant}}\\
\caption{{Average of sum spectral efficiency using the MMF DL power control and different UL pilot power control strategies, different number of multicast subgroups, and different number of BS antennas. $M=64$, $P_{\mathrm{dl}}=33$ dBm, $Q_{\mathrm{ul}}=20$ dBm, 5 clusters, 8 users per cluster.}}
\label{fig:number_antennas}
\end{figure*}

Our final evaluation focuses on the effects of the number of \gls{bs} antennas on the performance of the proposed subgrouping and UL pilot power control strategies. 
Figure \ref{fig:number_antennas} illustrates the effects of the number of \gls{bs} antennas in the \gls{sse} achieved by $40$ users deployed in $5$ clusters. As expected, we observe that increasing the number of antennas at the BS yields a higher average \gls{sse}, as a result of a better spatial resolution of the multicast transmissions as well as an increase of the favorable propagation conditions. Optimal solution of using $5$ subgroups and UL pilot power control based on Algorithm 1 provides an average \gls{sse} of $16.64$, $27.47$, and $34.8$ b/s/Hz for $64$, $128$, and $192$, BS antennas, respectively. Again, we notice that regardless of the number of BS antennas, using the optimal subgroup setup along with the UL pilot power control of Algorithm 1 provides the highest \gls{sse} performance. 

\section{Conclusion and Future Perspectives}
\label{sec:conclusion}
This work considers spatially correlated Rayleigh fading channels in a single-cell massive \gls{mimo} system underlying multicast communications. We propose a multicast user subgrouping strategy which capitalizes on the knowledge of the users' spatial channel correlation characteristics to set up multiple multicast transmissions, one per each multicast subgroup, almost free of mutual interference. This is attained by grouping together the users with low levels of channel mutual orthogonality. We also defined an optimal \gls{mmf} \gls{dl} power control strategy to maximize the minimum \gls{se} among the different multicast subgroups, and developed a heuristic UL \gls{mmf} pilot power control strategy to mitigate the intra-subgroup interference and pilot contamination.

Our exhaustive simulation campaign reveals the benefits, in terms of \gls{sse}, provided by our subgrouping and power allocation strategies under different system setups, comparing that with alternative subgrouping and UL pilot power control strategies, and considering different precoding techniques. 
We also evaluated the effects on the performance of varying the number of users, the number of users per cluster, the cluster radius, the UL pilot and DL power budget, and the number of BS antennas. 
Tha main conclusion is that our proposed subgroup multicast transmission strategy along with the proposed UL/DL power control allocations always yields the best performance. Importantly, whenever the optimal number of subgroups is selected, a more efficient utilization of the radio resources is ensured, and using low-complexity precoding and UL power allocation schemes leads anyhow to excellent values of \gls{sse}. The optimal number of subgroups can be easily identified by inspecting the users' location clusters, capitalizing on the knowledge on the channel statistics. 
We demonstrated that grouping the users by their channel spatial similarities significantly improves the effectiveness of the multicast service. However, the number of location clusters was an input simulation parameter, hence a known value. How identifying the number of clusters given a users deployment, and thereby obtaining the optimal number of multicast subgroups was out of the scope of this work. Hence, an interesting future research would include investigating deep learning techniques to find the optimal number of multicast subgroups based on a priori knoweledge of the users' covariance matrices. 

Massive MIMO technology builds upon spatial division multiple access (SDMA), and thanks to its aggressive spatial multiplexing of the users, enables a scalable, efficient implementation of multi-user communications. In general, multicasting benefits from Massive MIMO technology and, as demonstrated by this work, multiplexing multicast subgroups of users in the spatial domain over non-orthogonal time-frequency resources further increases the SE and decreases the estimation overhead. Alternative or complementary solutions to SDMA may include power-domain non-orthogonal multiple access (NOMA). NOMA, which has been recently promoted as a solution for 5G and beyond, consists in multiplexing the users in the power domain, entailing the need of removing the multi-user interference at the receiver by means of successive interference cancellation (SIC) techniques to separate the superimposed signals, hence requiring a higher complexity decoding. NOMA has been shown to provide higher SE and system capacity than OMA in the downlink of single-antenna systems \cite{Vaezi2019}, but fails in fully exploiting the spatial degrees of freedom in multi-antenna setups, as demonstrated in \cite{Clerckx2021}.
NOMA has been also considered as a booster of multicast device-to-device (MD2D) communications, wherein users in proximity are allowed to communicate directly over a D2D link to exchange multicast content. This solution is efficient in terms of sum rate but poorly performs in fairness \cite{Hmila2021}, hence unsuitable for our subgrouping framework relying on max-min fairness optimization. 
Rate-Splitting multiple access (RSMA) embodies the advantages of both SDMA and NOMA, as it fully utilizes the spatial degrees of freedom and fully decodes the multi-user interference \cite{Mao2018}. It constitutes an appealing research direction especially for subgrouping-based multicasting.

Last but not least, a further extension to this work may include the analysis of the proposed multicast subgrouping strategies in distributed cell-free massive MIMO systems~\cite{Interdonato2019}.


\bibliographystyle{IEEEtran}
\bibliography{main}
\end{document}